\documentclass[10pt, conference, compsocconf]{IEEEtran}

\usepackage{times}
\usepackage{array,booktabs,amsmath,graphicx,fancyvrb,tabularx,longtable}
\usepackage{subfig}
\usepackage{url}

\let\labelindent\relax

\usepackage{enumitem}
\usepackage{multirow}
\usepackage{balance}

\begin{document}

\renewcommand\ttdefault{cmvtt}

\title{Scission: Performance-driven and Context-aware Cloud-Edge\\Distribution of Deep Neural Networks}

\author{\IEEEauthorblockN{Luke Lockhart\IEEEauthorrefmark{1},
Paul Harvey\IEEEauthorrefmark{2},
Pierre Imai\IEEEauthorrefmark{2}, 
Peter Willis\IEEEauthorrefmark{3} and
Blesson Varghese\IEEEauthorrefmark{1}}
\IEEEauthorblockA{\IEEEauthorrefmark{1}Queen's University Belfast, UK; Email: \{llockhart04, b.varghese\}@qub.ac.uk}
\IEEEauthorblockA{\IEEEauthorrefmark{2}Rakuten Mobile, Japan; Email: \{paul.harvey, pierre.imai\}@rakuten.com}
\IEEEauthorblockA{\IEEEauthorrefmark{3}British Telecommunications plc, UK; Email: peter.j.willis@bt.com}
}

\maketitle

\begin{abstract}
Partitioning and distributing deep neural networks (DNNs) across end-devices, edge resources and the cloud has a potential twofold advantage: preserving privacy of the input data, and reducing the ingress bandwidth demand beyond the edge. However, for a given DNN, identifying the optimal partition configuration for distributing the DNN that maximizes performance is a significant challenge. This is because the combination of potential target hardware resources that maximizes performance and the sequence of layers of the DNN that should be distributed across the target resources needs to be determined, while accounting for user-defined objectives/constraints for partitioning. This paper presents \texttt{Scission}, a tool for automated benchmarking of DNNs on a given set of target device, edge and cloud resources for determining optimal partitions that maximize DNN performance. The decision-making approach is context-aware by capitalizing on hardware capabilities of the target resources, their locality, the characteristics of DNN layers, and the network condition. Experimental studies are carried out on 18 DNNs. The decisions made by \texttt{Scission} cannot be manually made by a human given the complexity and the number of dimensions affecting the search space. The benchmarking overheads of \texttt{Scission} allow for responding to operational changes periodically rather than in real-time. \texttt{Scission} is available for public download\footnote{\url{https://github.com/qub-blesson/Scission}}. 


\end{abstract}


\IEEEpeerreviewmaketitle

\section{Introduction}
\label{sec:introduction}
Deep Neural Networks (DNNs) are integral to image, video or speech recognition applications~\cite{dnn-01,dnn-02}. 
A DNN is a sequence of multiple layer types, such as convolution, activation or pooling, that have varying computational requirements. The output size of each layer depends on the layer type and configuration.

\begin{figure*}[ht]
	\centering
	\includegraphics[width=0.99\textwidth]{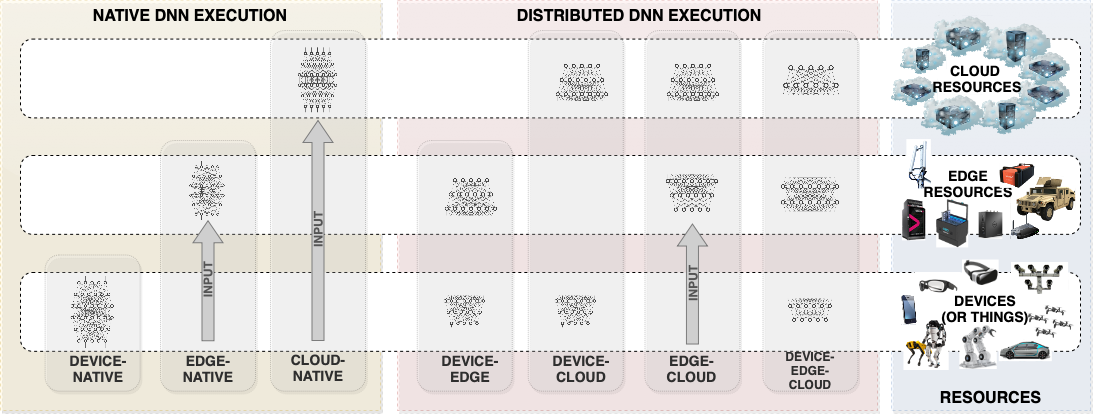}
	\caption{Native and distributed DNN execution options on a three-tier resource pipeline (devices may also be called `things').}
\label{fig:dnn-native-distributed}
\end{figure*}


Recently, distributed execution of the DNN across the cloud and resources at the edge of the network, within the edge computing paradigm~\cite{edgecomp-01, edgecomp-02, edgecomp-03}, has been found to be beneficial~\cite{edge-dnn-survey-1, edge-dnn-survey-2}. The advantages offered by using the edge are privacy preservation, reduced ingress bandwidth demand, and reduced inference times. 
A distributed execution approach could either execute the entire DNN on the edge if there are sufficient compute resources available or act as a pre-filter (partially processed) for the input data before it is sent in the WAN to the cloud. The edge can also be an aggregation point in use-cases, for example, a network of drones or cameras that are linked to an edge resource.
It has also been demonstrated that for data streams, the frame drop rate can be reduced at the edge when compared to the cloud~\cite{couper}.   
Additionally, resources at the edge may be powered by the main lines and may have relatively more compute capabilities than the end device, thereby providing opportunities for executing large DNNs while being sufficiently accurate. 


Leveraging the edge provides numerous possibilities for distributing a DNN in addition to those when only using the device and the cloud~\cite{edge-dnn-survey-1, edge-dnn-survey-2}. These possibilities as shown in Figure~\ref{fig:dnn-native-distributed} are: (i) edge-native execution of the DNN, (ii) distributed execution across the edge and the device, or (iii) distributed execution across the cloud, edge and device. 
In edge-native execution, all the layers of the DNN will run on the edge and in distributed execution, a specific sequence of layers will run on each resource. 
However, \textit{for any given DNN, identifying the execution approach that maximizes its performance} is not a trivial challenge. This is because the following three associated questions need to be addressed:

(\textbf{Q1}) \textit{Which combination of potential target hardware resources maximizes performance?} This question requires the identification of whether native or distributed execution approaches are best suited for a given DNN on a set of resources, comprising the device, edge, and cloud. Also, if there are multiple device, edge or cloud choices, which target resource(s) should be selected for deploying the DNN.

(\textbf{Q2}) \textit{Which sequence of layers should be distributed across the target resource(s) for maximizing DNN performance?} DNNs can have a large number of layers with varying computational requirements and output sizes. 
For distributed execution, the layers at which a DNN is partitioned for optimal performance needs to be identified.
This cannot be done manually because there are DNNs that could have a large number of layers. For example, DNNs such as NASNetLarge has 1041 layers and InceptionResNetv2 has 782 layers. In addition, DNNs cannot be partitioned at all layers (will be discussed in Section~\ref{sec:scission}). An ad hoc distribution of a DNN that arbitrarily selects the sequence of layers would result in under-performing DNNs.

(\textbf{Q3}) \textit{How can the performance of DNNs be optimized given user-defined objectives or constraints?} Although addressing Q1 and Q2 will provide an ideal partition of a DNN for a given set of hardware resources, they may not be optimal when user-defined objectives or constraints are taken into account. For example, although a cloud native execution approach may be ideal for maximizing the performance of a DNN, an application owner may want to run a specific sequence of layers on the edge for enhancing data privacy or reducing the volume of output data sent to the cloud. If an edge resource has to undergo maintenance, then an administrator may require the DNN to be redistributed across the cloud and the device, which would need a different partition configuration.

To address the above challenge and associated questions, this paper makes the following \textit{research contributions}: 

(1) Proposes \texttt{Scission}, a tool for automated benchmarking of DNNs on a given set of target device, edge and cloud resources for determining the optimal partition for maximizing DNN performance. 

(2) Develops the underpinning benchmarking approach of \texttt{Scission} that collects benchmark data by executing the DNNs on all target resources and subsequently identifies whether a native or distributed execution approach is most suited for the DNN (addresses Q1). For distributed execution, it identifies the optimal resource pipeline and partitions measured by the lowest end-to-end latency (compute time on resources and the communication time between resources) of the DNN by: (a) pairing the most computationally intensive layers with capable resources to minimize compute latencies, and at the same time (b) selecting layers with the least amount of output data as potential end layers of a partition to minimize communication latencies (addresses Q2). Thus the decision-making approach in \texttt{Scission} is context-aware by capitalizing on the hardware capabilities of the target resources, their locality, the characteristics of DNN layers, and network condition. 

(3) Provides a querying engine that has less than a 50 millisecond overhead to ensure that user-define constraints or objectives can be taken into account for determining optimal partitions that maximize the performance of distributed DNNs (addresses Q3). 


(4) An experimental study to demonstrate that \texttt{Scission} can facilitate: (a) DNN partitioning under different network conditions, (b) DNN partitioning under different input data sizes, (c) DNN partitioning under user-defined constraints, (d) DNN partitioning for comparing different target resource pipelines, and (e) the identification of the top $N$ DNN partitions that maximize performance.
It is observed that ideal DNN partitioning needs to be context and data-driven and it is impossible to determine optimal partitions manually. \texttt{Scission} achieves these and is a valuable tool for deploying context-aware and distributed DNNs in an cloud-edge environment.

The remainder of this paper is organized as follows. 
Section~\ref{sec:scission} provides a background to the DNN models considered in this paper and presents an overview of the underpinning methodology for benchmarking, decision-making and querying of \texttt{Scission}. 
Section~\ref{sec:experimentalstudies} presents the results obtained from an experimental study on \texttt{Scission}. 
Section~\ref{sec:relatedwork} presents related work. 
Section~\ref{sec:conclusions} concludes this paper by presenting avenues for future research.

\section{Scission}
\label{sec:scission}
This section firstly provides a background to DNNs and the types of DNNs that are considered in this paper, which is followed by observations that led to the development of \texttt{Scission}. The architecture, the underlying benchmarking approach, the context-aware decision-making process, and finally the querying capability of \texttt{Scission} are then presented. 

\subsection{Background}
\label{sec:background}
A DNN is a sequence of layers and is a general term that covers all neural networks with multiple hidden layers (that is multiple layers between the input and output layers)~\cite{dnn-01,dnn-02}. A DNN may consist of different layers and the most common types are as follows: 
1)~Fully-connected layers connect every neuron to all neurons in the previous layer with the aim of preforming high-level reasoning.
2)~Convolution layers convolve the input to produce feature maps of inputs with the aim of learning features.
3)~Pooling layers apply a pre-defined function (maximum or average) to down sample the input.
4)~Activation layers apply non-linear functions and the most commonly used is the rectified linear unit (ReLu).
5)~A Softmax layer is generally used for classification with the aim of generating a probability distribution over the possible classes.

\begin{table}[t]
\centering

\footnotesize

\caption{Pre-trained DNN models from Keras used in this paper; Type: L - linear, B - branching}
\label{tab:pretrainedmodels}
\begin{tabular}{@{}lcccc@{}}

\hline
\textbf{DNN Model}             & \textbf{Size (MB)}  & \textbf{Layers} & \textbf{Partition points} & \textbf{Type}\\
\hline
Xception \cite{xception}        & 88  & 134    & 13           & B\\
VGG16 \cite{vgg}             & 528 & 23     & 21           & L\\
VGG19 \cite{vgg}            & 549  & 26     & 24           & L\\
ResNet50 \cite{resnet}          & 98   & 177    & 23           & B\\
ResNet101 \cite{resnet}        & 171  & 347    & 40           & B\\
ResNet152 \cite{resnet}       & 232  & 517    & 57           & B\\
ResNet50V2 \cite{resnet}       & 98   & 192    & 15           & B\\
ResNet101V2 \cite{resnet}      & 171 & 379    & 15           & B\\
ResNet152V2 \cite{resnet}      & 232 & 556    & 15           & B\\
InceptionV3 \cite{inceptionv3}      & 92  & 313    & 18           & B\\
InceptionResNetV2 \cite{inceptionResnetv2} & 215 & 782    & 60           & B\\
MobileNet \cite{mobilenet}       & 16  & 93     & 91           & L\\
MobileNetV2 \cite{mobilenetv2}      & 14  & 157    & 65           & B\\
DenseNet121 \cite{densenet}      & 33  & 429    & 21           & B\\
DenseNet169 \cite{densenet}      & 57  & 597    & 21           & B\\
DenseNet201 \cite{densenet}     & 80  & 709    & 21           & B\\
NASNetMobile \cite{nasnet}     & 23  & 771    & 4            & B\\
NASNetLarge \cite{nasnet}       & 343 & 1041   & 4            & B\\ 

\hline
\end{tabular}
\end{table}

In this paper, 18 DNNs as shown in Table~\ref{tab:pretrainedmodels} are considered. The table presents the size of a trained model and its corresponding weights, the total number of layers in the DNN (including input and output layers), the number of valid points for partitioning, and the type of the DNN. These models are explored in the context of Keras\footnote{\url{https://keras.io}}, an open source neural network library that runs on TensorFlow\footnote{\url{https://www.tensorflow.org/}}. These models are trained on the ImageNet database~\cite{imagenet}.

\begin{figure}[t]
\begin{center}
	\subfloat[Linear DNN]
	{\label{fig:lineardnn}
	\includegraphics[width=0.49\textwidth]{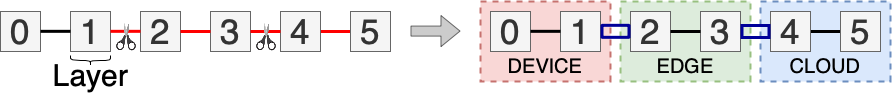}}
\hfill
	\subfloat[Branching DNN]
	{\label{fig:branchingdnn}
	\includegraphics[width=0.49\textwidth]{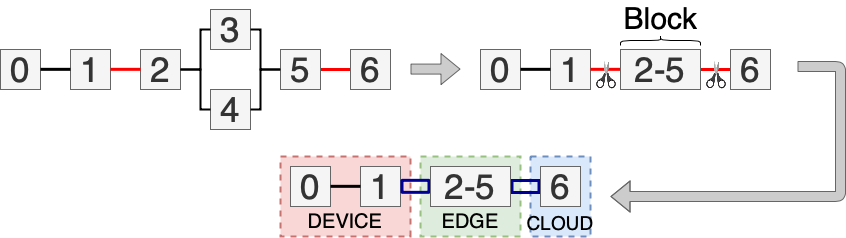}}
\end{center}
\caption{An example of partitioning a linear and branching DNN across the entire resource pipeline comprising device, edge and cloud. An example of a block is shown in Figure~\ref{fig:branchingdnn} (Layers 2-5). Red connectors between layers/blocks are valid partitioning points. Blue connectors are inter-resource communication.}
\label{fig:dnnpartitioning}
\end{figure} 

\begin{figure}[t]
	\centering
	\includegraphics[width=0.49\textwidth]{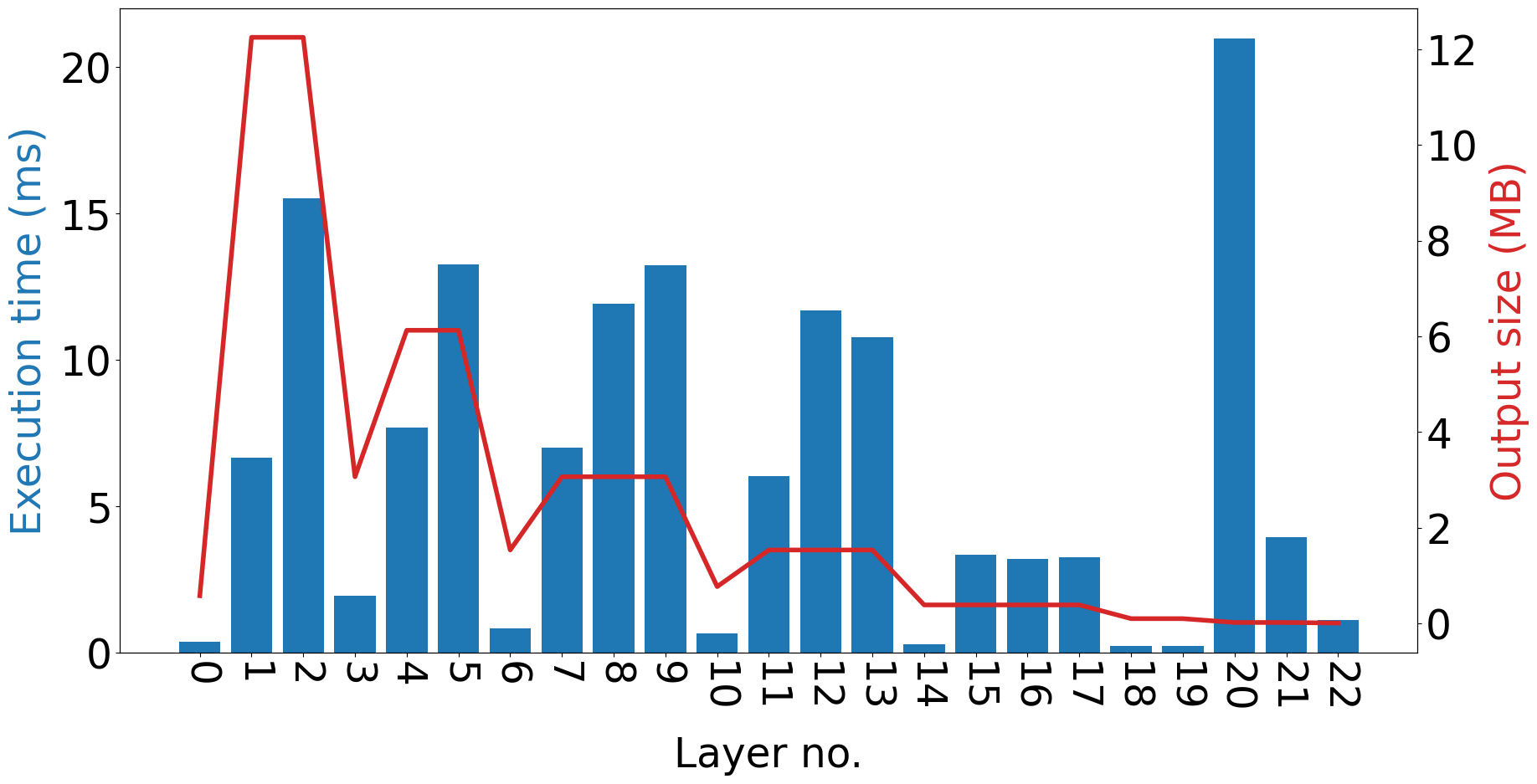}
	\caption{Average execution time (of five runs) and output data size of each layer of VGG16 on a `Cloud' resource (refer Table~\ref{tab:hardware}).}
\label{fig:vgg16-cloud}
\end{figure}

Two categories of DNNs are considered, namely linear and branching. 
In a linear DNN, the neural network is sequential - the input of one layer is connected to the next. This results in a singular path between the first and last layers as seen in Figure~\ref{fig:lineardnn}. 
Figure~\ref{fig:vgg16-cloud} shows the execution time and the output data size of the 23 different layers of VGG16, an example linear model (executed on the `Cloud' resource shown in Table~\ref{tab:hardware}). It is noted that the layers have varying execution time and the output sizes of the layers vary. 

\begin{figure}[t]
	\centering
	\includegraphics[width=0.49\textwidth]{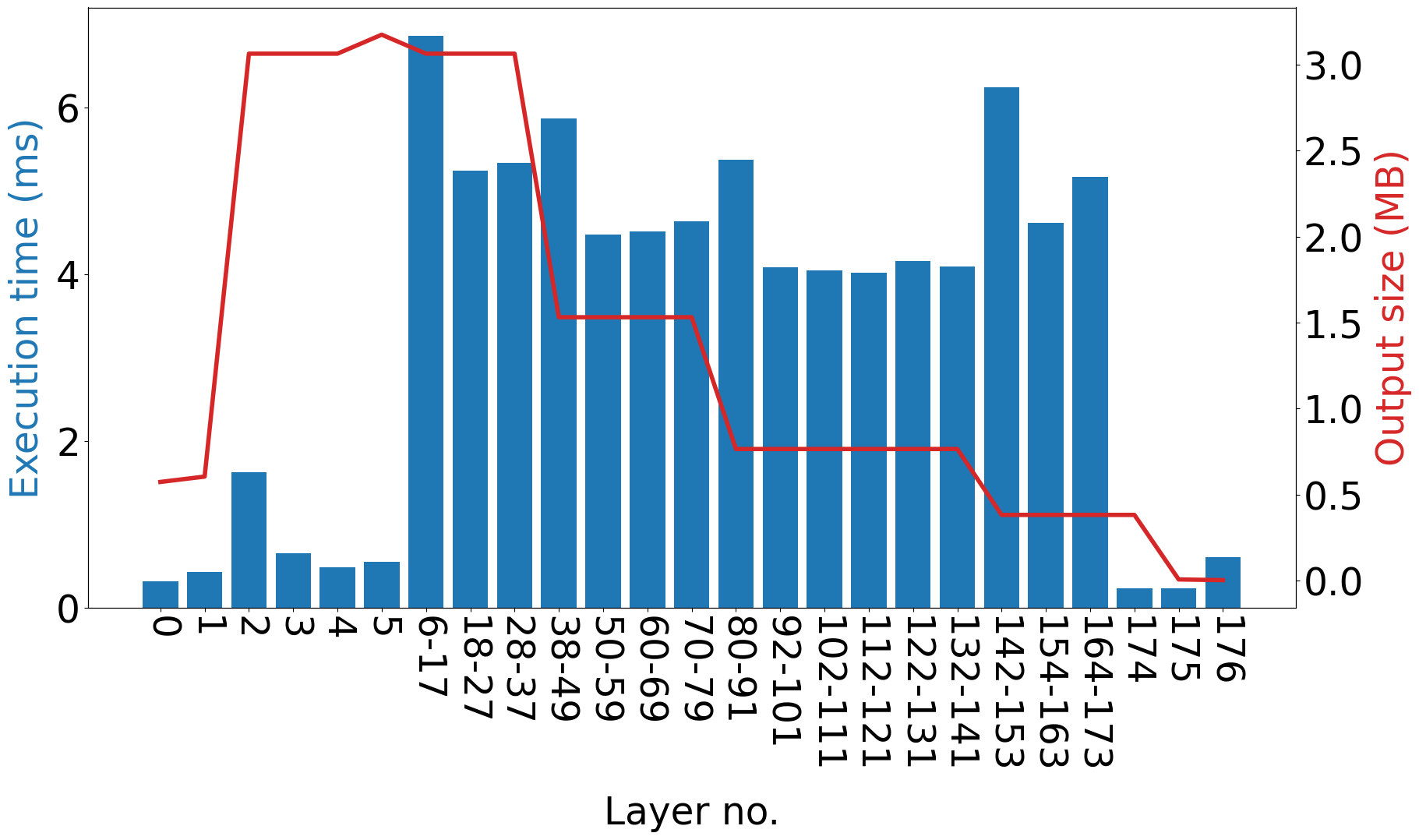}
	\caption{Average execution time (of five runs) and output data size of each layer of ResNet50 on a `Cloud' resource (refer Table~\ref{tab:hardware}).}
\label{fig:resnet50-cloud}
\end{figure}


If a linear DNN that has $N$ layers needs to be distributed across two resources, then a partitioning approach would need to create two partitions of the DNN. The first partition would consist of a sequence of the first $x$ layers and the second partition would consist of $N-x$ layers. The output of the $x^{th}$ layer would need to be provided as an input layer for the second partition. 
DNNs naturally lend themselves to distributed execution as their segmented structure provides rational points to partition. 
There are $N-2$ potential partitioning points (rather than $N-1$) since a partition configuration in which the first partition comprises only the first layer, results in the second partition containing a duplicated input layer.
Figure~\ref{fig:lineardnn} provides an example of a linear DNN that is distributed across a resource pipeline consisting of the device, edge, and cloud. The red connectors show the valid partitioning points in the linear model.

On the other hand, in a branching DNN, a layer may be connected to more than two layers which results in parallel paths between the first and last layers. 
Partitioning a model in a parallel region can lead to synchronization issues and may add additional communication overhead as multiple metadata outputs will need to be transferred from one resource to another~\cite{couper}. Therefore, layers within a branch are grouped together as a block of layers and treated as a single entity. This reduces the number of partitioning points (for example, the ResNet50 DNN has 177 layers, but only 23 valid partition points as shown in Table~\ref{tab:pretrainedmodels}). Figure~\ref{fig:resnet50-cloud} shows the execution time and the output data size of the 25 different entities (layers and blocks are identified by the layer numbers) of ResNet50, an example branching model. Again the variable execution time and output data size of layers/blocks are noted. 
Figure~\ref{fig:branchingdnn} shows an example of a branching DNN that is distributed across the device, edge and cloud. Layers 2-5 are considered as a single block.

The above highlights that DNNs may have a large number of layers and may take the form of linear or branching models. The execution time of individual layers and the output size vary for each layer. 
If a DNN needs to be distributed across multiple resources, it would be impossible to manually determine efficient partition configurations. This is due to the potentially complex structure of a DNN and a large search space arising from the combination of partitioning points, target hardware resources, and optimization criteria. Therefore, an automated approach for DNN partitioning is required.

\subsection{Motivation}
\label{subsec:motivation}

\begin{figure*}[ht]
	\centering
	\includegraphics[width=0.95\textwidth]{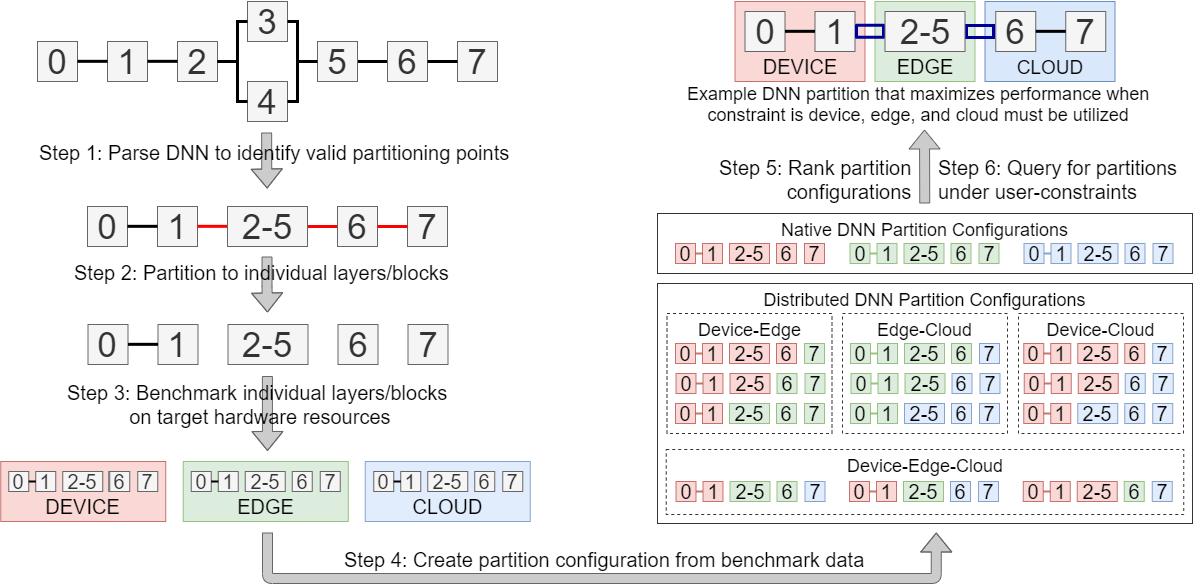}
	\caption{The underlying six-step benchmarking and partitioning methodology of \texttt{Scission}.}
\label{fig:methodology}
\end{figure*}

\texttt{Scisssion} proposed in this paper is designed on the following six practical observations to make it widely applicable for maximizing the performance of DNNs: 

(i) \textit{DNN partitioning must account for multiple resource tiers in cloud-edge continuum}. Many options for distributing large DNNs become available as more resource tiers between the cloud-edge continuum become accessible for computing. The approach for identifying optimal partitions of DNNs should scale across resource tiers. This paper considers the device, edge and cloud tiers.  

(ii) \textit{DNN partitioning must be based on empirical data obtained from the underlying hardware rather than based on estimates}. A large body of existing research estimates optimal partitions by relying on predicted performance on a given resource by making assumptions of the target hardware platform. However, modern hardware is known to have complex processor and memory architectures that sometimes results in a non-linear relationship between performance and the amount of resource~\cite{motivation-01}. Hence, empirical data based partitioning will be more reliable than alternate approaches. 

(iii) \textit{DNN partitioning must be able to identify a set of performance efficient partitions}. This is important because the most efficient DNN partitions may only have a negligible improvement over the other partitions, which may be more practical due to organizational or geo-political reasons.  

(iv) \textit{DNN partitioning needs to be context-aware across multiple dimensions}. Identifying performance efficient partitions is not only dependent on DNN layer characteristics and output data. Performance is affected by hardware capabilities of the target platform, resource locality, load and failures, and network condition between resources. These dimensions need to be taken into account. 

(v) \textit{DNN partitioning must account for user-defined objectives or constraints}. A performance efficient partition obtained by optimizing against the dimensions described above may not always be ideal. A human must be able to specify constraints as input to the partitioning process. For example, an application administrator may want a particular sequence of layers to be executed on an end device for retaining intermediate data of a few layers on the device although it affects the overall end-to-end latency. 

(vi) \textit{Practical DNN partitioning must be rapid}. Variations in network conditions and changes to resource workloads may affect the optimal partition points of a DNN. For example, the available bandwidth to a drone may increase as it navigates away from a low coverage area. This may result in the DNN to be partitioned from device-native (which may be less energy efficient) to be distributed across the device-edge-cloud. This repartitioning needs to occur with low overheads to be advantageous in real-world use (the worst case prediction for \texttt{Scission} only takes 0.05 seconds on the cloud).

\subsection{Partition Methodology}
The six step methodology for automated partitioning adopted by \texttt{Scission} is shown in Figure~\ref{fig:methodology} and described below:

\textit{Step~1: Parse the DNN to find valid partitioning points}. As presented in Section~\ref{sec:background} the DNN is parsed to identify valid partitioning points. For a linear DNN this is straightforward, where as for a branching DNN, the parallel paths need to be identified. Layers within the parallel path are considered as a single entity, referred to as block. As shown in Figure~\ref{fig:methodology} the red connectors show the valid partitioning points. 

\textit{Step~2: Partition into individual layers/blocks}. This step ensures that the DNN is partitioned into distinct sub-models with individual layers or blocks for the purposes of benchmarking. It should be noted that each sub models requires an input layer to facilitate the processing of the output from the previous layer.

\textit{Step~3: Benchmark each layer/block on target hardware resources}. In this step, given a set of target hardware resources, such as the device, edge, or cloud, each layer/block is benchmarked five times. The average execution time and the output data size is recorded. The 18 DNNs shown in Table~\ref{tab:pretrainedmodels} are considered in this paper. 

\textit{Step~4: Create partition configurations from benchmark data}. 
The benchmark data comprises the average execution time of each layer/block. 
The communication overhead to transfer output data across different resources is calculated from user-provided data, such as the average bandwidth available. 
This data is used to exhaustively develop partition configurations such that the end-to-end latency (compute and communication overheads) of all combinations of layers/blocks paired to different resources are known.  

Two types of partition configurations are considered by \texttt{Scission}, namely native and distributed as shown in Figure~\ref{fig:dnn-native-distributed}. Native partition configurations are those in which all layers/blocks execute on a single resource (for example, device-native, edge-native, or cloud-native). Distributed partition configuration are those in which the DNN collaborates across multiple resources by executing the layers/blocks on multiple resources (for example, distributed execution across device-edge, device-cloud, and device-edge-cloud).  

\textit{Step~5: Rank partition configurations}. 
Once all partition configurations have been generated, they are ranked. The ranking may be generated by optimizing against end-to-end latency (additional objectives, such as minimum data transfer across resources, or a combination of these can be provided in Step~6). The Top $N$ partition configurations are presented to the user. 

\textit{Step~6: Query \texttt{Scission} for partition configurations given user-defined constraints}.
\texttt{Scission} interacts with the user by not only providing the default rankings produced in Step~5, but also accepting user-defined constraints provided as queries. The example shown in Figure~\ref{fig:methodology} is the result of executing the query for the fastest DNN partition configuration that collaborates between all (device, edge and cloud) resources. Queries are not limited to only minimizing execution latency or bandwidth. For example, they may be constructed to:
\begin{itemize}[leftmargin=0.35cm]
    \item Apply bandwidth constraints (for example, the edge resource must not transfer more than 1MB to the cloud).
    \item Apply execution time constraints (for example, the execution time on the device must not exceed 1 second, or 30\% of the overall execution time must be on the edge).
    \item Include or exclude resources (for example, distribution must not include the cloud, or execution must be edge-native).
    \item Specify layer/block execution locations (for example, Layer 7 must execute on the edge).
\end{itemize}
The Top $N$ partition configurations are presented to the user. More complex queries can be provided to \texttt{Scission}. Examples include: (i) Find the partition configuration that results in the lowest execution latency, but the device and edge must not transfer more than 1MB. (ii) Find partition configuration that has the lowest inter-resource data transfer, but $n$ layers are executed on the edge. (iii) Find partition configuration with lowest end-to-end latency and does not use the cloud and at least half of the layers/blocks must be executed on the device.

\section{Experimental Studies}
\label{sec:experimentalstudies}
This section presents the experimental test bed and software set up and is followed by the results obtained from \texttt{Scission}.

\begin{table*}
\centering
\caption{Specification of the target hardware resources used}
\label{tab:hardware}
\begin{tabular}{@{}lcccccc@{}}
\hline
\textbf{Resource}       & 
\begin{tabular}[c]{@{}l@{}} \textbf{CPU} \\ \textbf{Arch.}\end{tabular} &
\begin{tabular}[c]{@{}l@{}} \textbf{CPU freq.} \\ \textbf{(GHz)}\end{tabular} & \begin{tabular}[c]{@{}l@{}}\textbf{CPU} \\ \textbf{cores}\end{tabular} & \begin{tabular}[c]{@{}l@{}}\textbf{RAM}\\ \textbf{(GB)}\end{tabular} & \textbf{GPU} &
\textbf{OS}\\ 
\hline
Device & ARMv8 &
1.5                                                       & 4                                                        & 4                                                  & N/A   & Raspbian Buster          \\
Edge (1)   & AMD64 & 4.5                                                       & 2                                                        & 4                                                  & N/A        & Ubuntu 18.04 LTS     \\
Edge (2)  & AMD64 & 3.7                                                       & 4                                                        & 8                                                  & N/A         & Ubuntu 18.04 LTS    \\
Cloud  & AMD64 & 4.5                                                       & 8                                                        & 32                                                 & N/A & Ubuntu 18.04 LTS\\
Cloud (with GPU) & AMD64 & 4.5                                                       & 8                                                        & 32                                                 & Nvidia GTX 1070 & Ubuntu 18.04 LTS \\ 
\hline
\end{tabular}
\end{table*}

\subsection{Setup}
\label{subsec:setup}

Experiments are carried out on hardware resources shown in Table~\ref{tab:hardware} to reflect a range of resources typically used. Two edge resources are employed with different hardware characteristics. Two cloud resources are used with and without a GPU. 

To emulate real world network performance, \texttt{Scission} uses the average network latency and bandwidth for: (i) 3G (1.6 Mbps upload and 67ms network latency)\footnote{\label{ofcommobile}\url{https://bit.ly/3hrGk4N}}, (ii) 4G (12.4Mbps upload and 55ms network latency)\textsuperscript{\ref{ofcommobile}}, and (iii) wired home fibre broadband (20Mbps upload and 20ms network latency)\footnote{\url{https://bit.ly/2EfHjqr}}. A network latency of 25ms and a bandwidth of 50Mbps is assumed for all edge-cloud connections. Results reported are average of five experimental runs. 

The \texttt{Scission} tool is implemented in Python and requires Tensorflow 2.0+ to be installed. Tensorflow is an end-to-end open source machine learning platform, which is used as the back end to run the pre-trained DNNs provided by Keras. NumPy is used for processing multi-dimensional arrays that are produced as layer outputs. 

\texttt{Scission} makes two assumptions. Firstly, the communication overheads can be calculated as $network$ $latency$ $+$ $data$ $size / bandwidth$. 
The second assumption is that the total inference time of a model is the sum of the execution times of individual layers or blocks. 
This assumption has been validated in previous research~\cite{assumption-02, dnntune}. 


\subsection{Results}
\label{subsec:results}

\begin{table*}[ht]
\centering
\caption{Overhead (in seconds) in benchmarking DNNs using the Scission partitioning methodology}
\label{tab:overhead}
\begin{tabular}{lccccc}
\hline
\textbf{DNN Model} & \textbf{Cloud} & \textbf{Cloud (with GPU)} & \textbf{Edge (1)} & \textbf{Edge (2)} & \textbf{Device} \\
\hline
Xception          & 2.95  & 2.11  & 7.07  & 6.78  & 36.10  \\
VGG16             & 3.85  & 1.93  & 8.52  & 9.83  & 44.88  \\
VGG19             & 2.98  & 1.92  & 8.43  & 11.86 & 51.82  \\
ResNet50          & 3.12  & 1.92  & 6.65  & 5.27  & 29.91  \\
ResNet101         & 6.03  & 5.19  & 12.42 & 9.55  & 57.67  \\
ResNet152V2       & 9.37  & 7.86  & 18.41 & 14.17 & 82.11  \\
ResNet50V2        & 3.31  & 2.77  & 6.59  & 5.27  & 33.30  \\
ResNet101V2       & 5.97  & 4.96  & 11.43 & 9.28  & 52.27  \\
ResNet152V2       & 8.85  & 7.49  & 17.27 & 14.23 & 77.99  \\
InceptionV3       & 4.93  & 4.36  & 9.46  & 7.48  & 43.46  \\
InceptionResNetV2 & 11.67 & 10.14 & 22.64 & 18.01 & 105.73 \\
MobileNet         & 1.66  & 1.52  & 3.60  & 2.97  & 14.09  \\
MobileNetV2       & 2.63  & 2.48  & 4.70  & 3.64  & 22.37  \\
DenseNet121       & 5.88  & 5.45  & 10.67 & 8.22  & 48.40  \\
DenseNet169       & 8.26  & 7.73  & 14.64 & 11.31 & 66.66  \\
DenseNet201       & 9.94  & 9.24  & 17.78 & 14.09 & 82.34  \\
NASNetMobile      & 10.40 & 9.95  & 17.42 & 13.02 & 77.45  \\
NASNetLarge       & 18.69 & 13.99 & 38.14 & 35.25 & 172.27 \\
\hline
\end{tabular}
\end{table*}

The experimental results obtained from \texttt{Scission} are exhaustive and discussing them entirely is outside the scope of this paper. However, the experiments and results to demonstrate the following five capabilities of \texttt{Scission} are considered in this paper: 1) DNN partitioning under different network conditions, 2) DNN partitioning under different input data sizes, 3) DNN partitioning under user-defined constraints, 4) DNN partitioning for comparing different target hardware resource pipelines, and 5) the top $N$ DNN partitions.
Sample results for executions on VGG19, ResNet50, MobileNetV2, InceptionV3 and DenseNet169 are presented. All experiments in this paper use a 150KB size input image unless otherwise stated.

The results from the above capabilities address \textbf{Q1}: `Which combination of potential target hardware resources maximizes performance?' that was posed in Section~\ref{sec:introduction}, but is specifically considered by the fourth capability. Similarly, all five capabilities will determine the best sequence of layers (or partition configuration) to address \textbf{Q2}: `Which sequence of layers should be distributed across the target platform for maximizing the DNN performance?' The third capability specifically addresses \textbf{Q3}: `How can the performance of DNNs be optimized given user-defined objectives or constraints?'

\begin{figure}[t]
\begin{center}
	\subfloat[3G]
	{\label{fig:3g-best-vgg19}
	\includegraphics[width=0.233\textwidth]{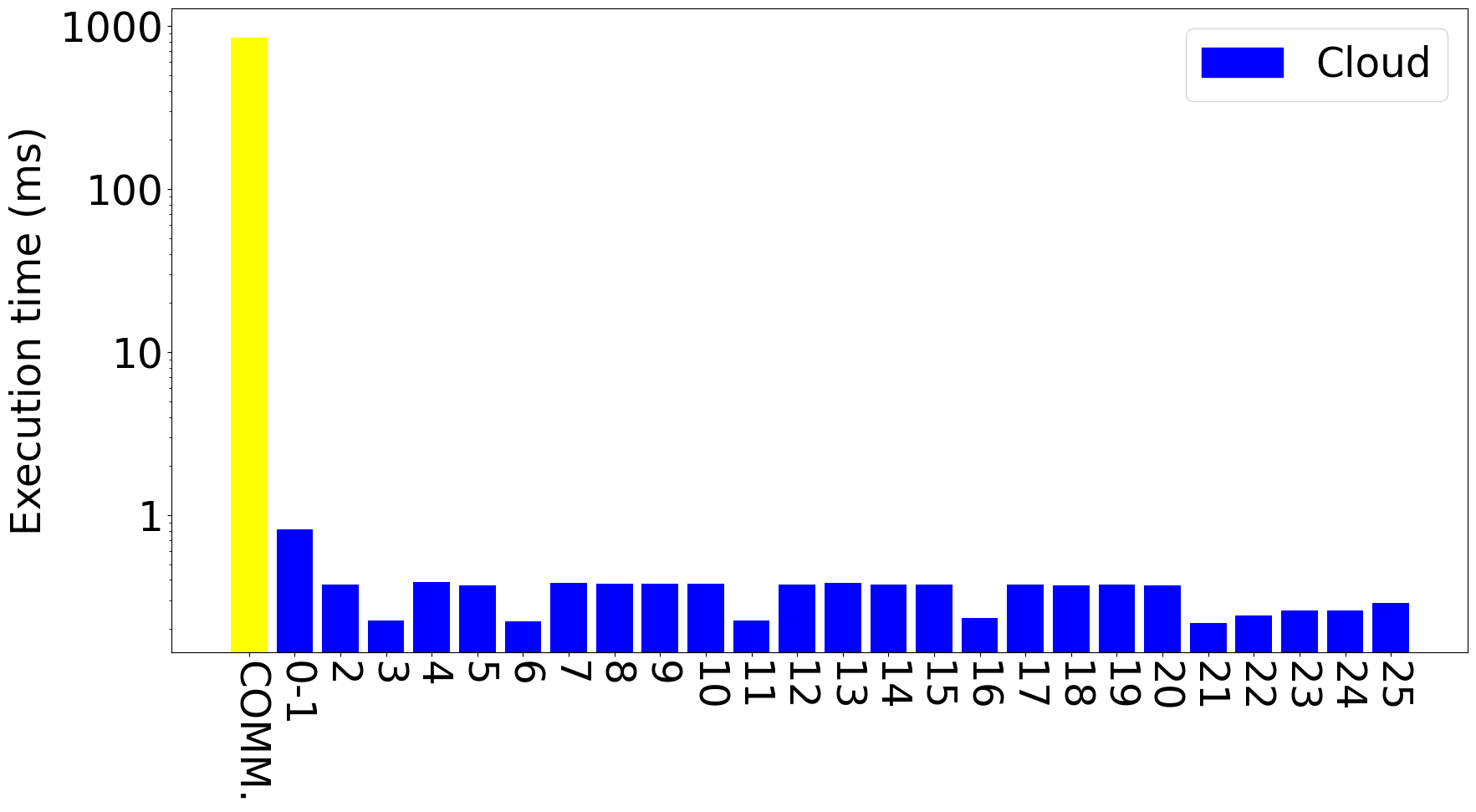}}
\hfill
	\subfloat[4G]
	{\label{fig:4g-best-vgg19}
	\includegraphics[width=0.233\textwidth]{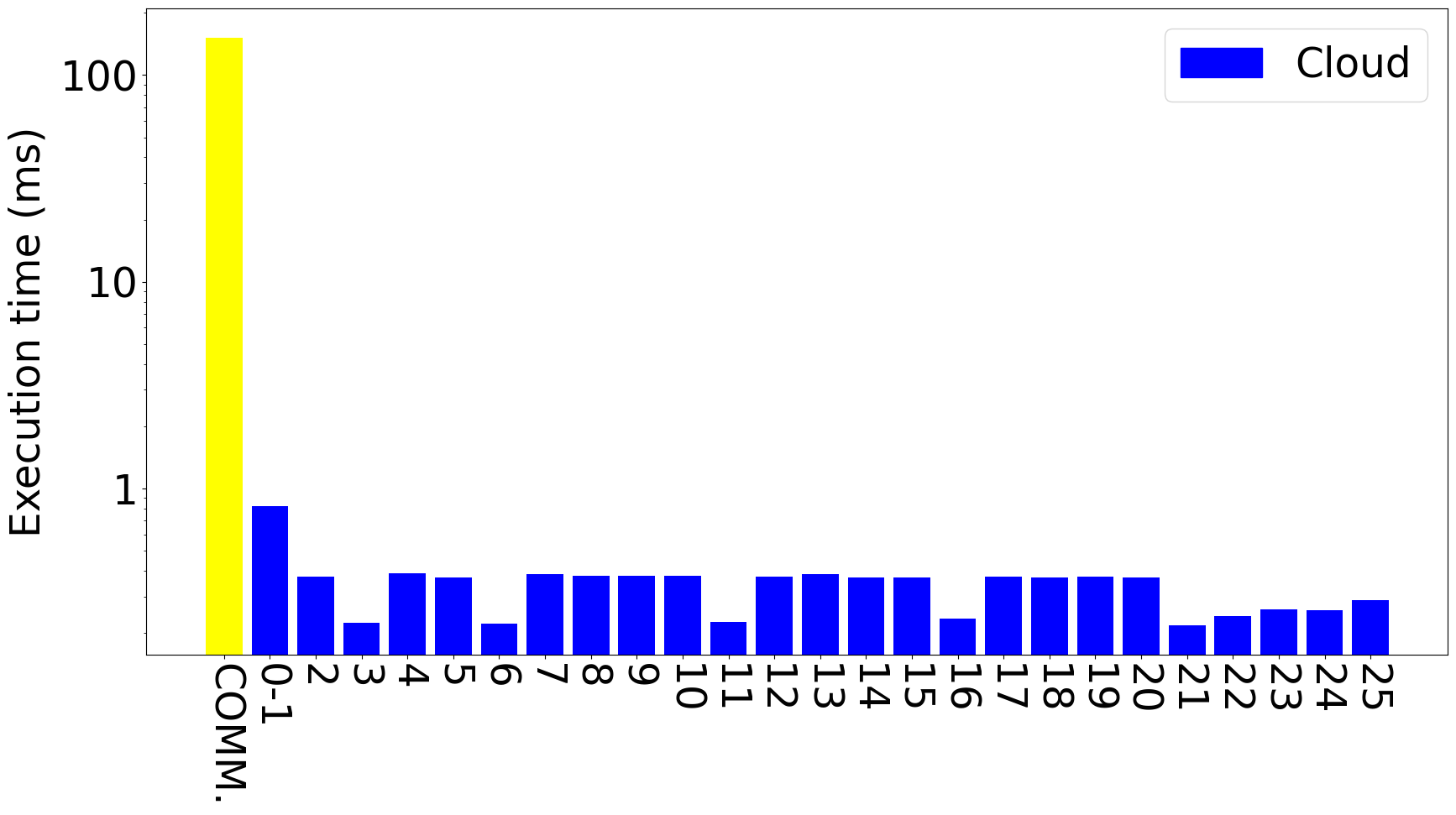}}
\end{center}
\caption{DNN partition of VGG19 with lowest end-to-end latency for different network conditions.}
\label{fig:network-vgg19}
\end{figure} 

\begin{figure}[t]
\begin{center}
	\subfloat[3G]
	{\label{fig:3g-best-resnet50}
	\includegraphics[width=0.233\textwidth]{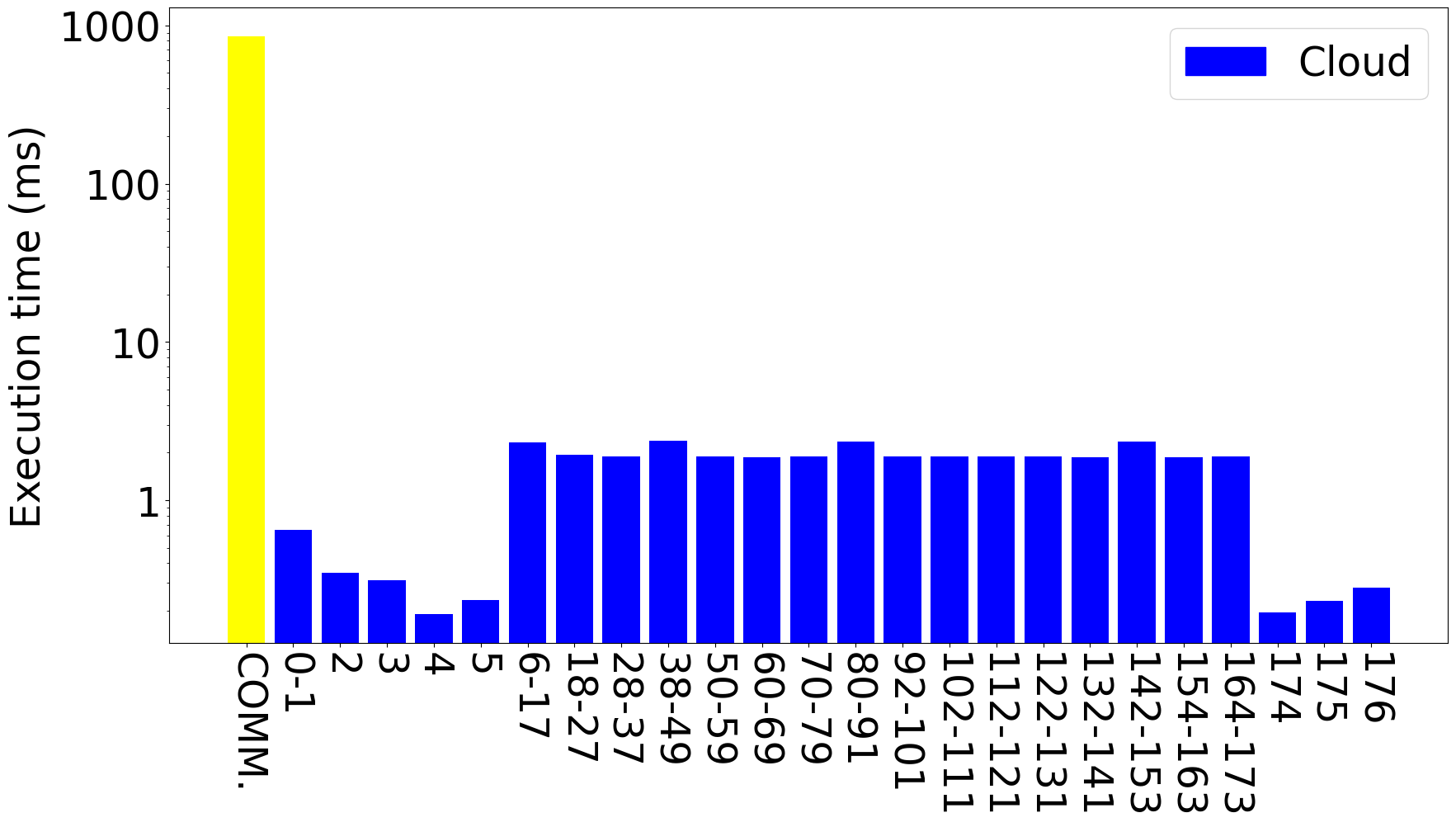}}
\hfill
	\subfloat[4G]
	{\label{fig:4g-best-resnet50}
	\includegraphics[width=0.233\textwidth]{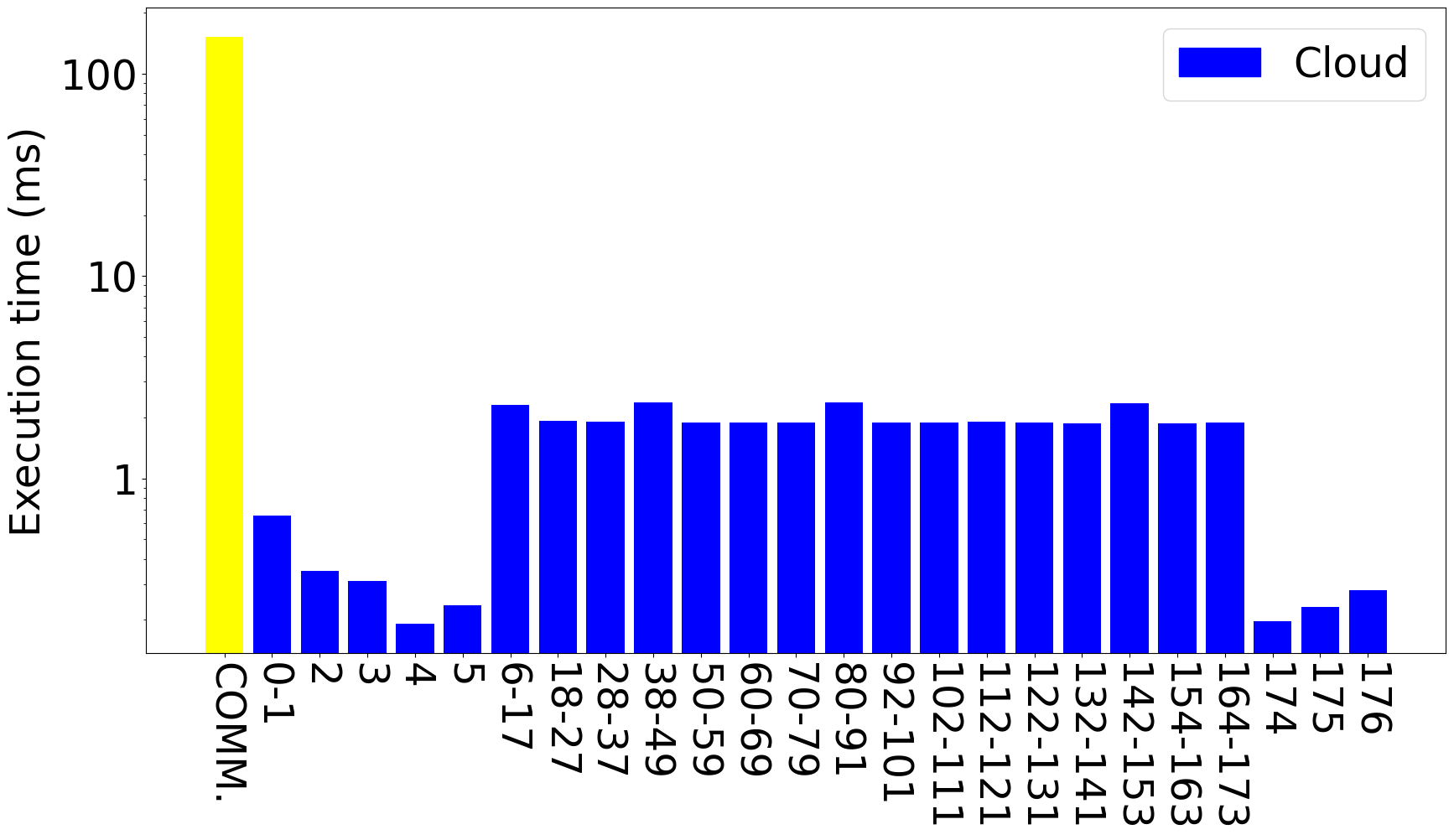}}
\end{center}
\caption{DNN partition of ResNet50 with lowest end-to-end latency for different network conditions.}
\label{fig:network-resnet50}
\end{figure} 

The time taken by the Scission partitioning methodology (overhead) is shown in Table~\ref{tab:overhead} using an input image of size 150KB. The time on the cloud, edge and device is proportional to the number of layers in the DNN model. Gathering benchmark data for the DNNs on the device takes the most time as expected. However, if a dedicated device were to be utilized, then the DNN will only need to be benchmarked once in an offline manner. The edge resources can benchmark all DNNs (except NASNetLarge) in under half a minute. Given these overheads, the methodology cannot be used in a highly transient environment, but can be used to respond to operational changes periodically. The querying time on the benchmark data is just under 50 milliseconds if two resource (cloud-edge-device) pipelines are considered. 



\subsubsection{DNN partitioning under different network conditions}
The results obtained from \texttt{Scission} highlight that DNN partitioning is affected by different network conditions (the optimal partitions for the same DNN may be different under different network conditions).

Figure~\ref{fig:network-vgg19} and Figure~\ref{fig:network-resnet50} show that the lowest end-to-end latency execution of VGG19 and ResNet50, respectively, under 3G and 4G conditions would be obtained if the DNN is cloud-native. This is because the cloud resource in terms of its execution performance is much faster than the device and edge resource utilized in this experiment. The communication overhead of 800ms of sending the image from the device to the cloud does not offset the compute performance obtained on the resource. 

\begin{figure}[ht]
\centering
\begin{center}
	\subfloat[3G]
	{\label{fig:3g-best-mobilenetv2}
	\includegraphics[width=0.5\textwidth]{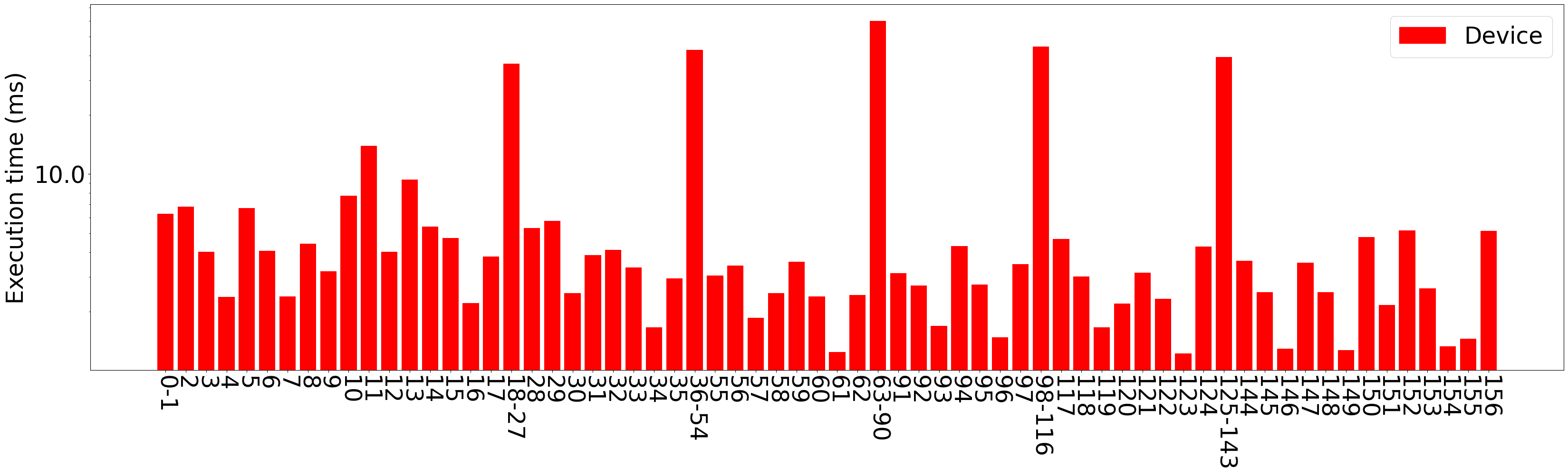}}
\hfill
	\subfloat[4G]
	{\label{fig:4g-best-mobilenetv2}
	\includegraphics[width=0.5\textwidth]{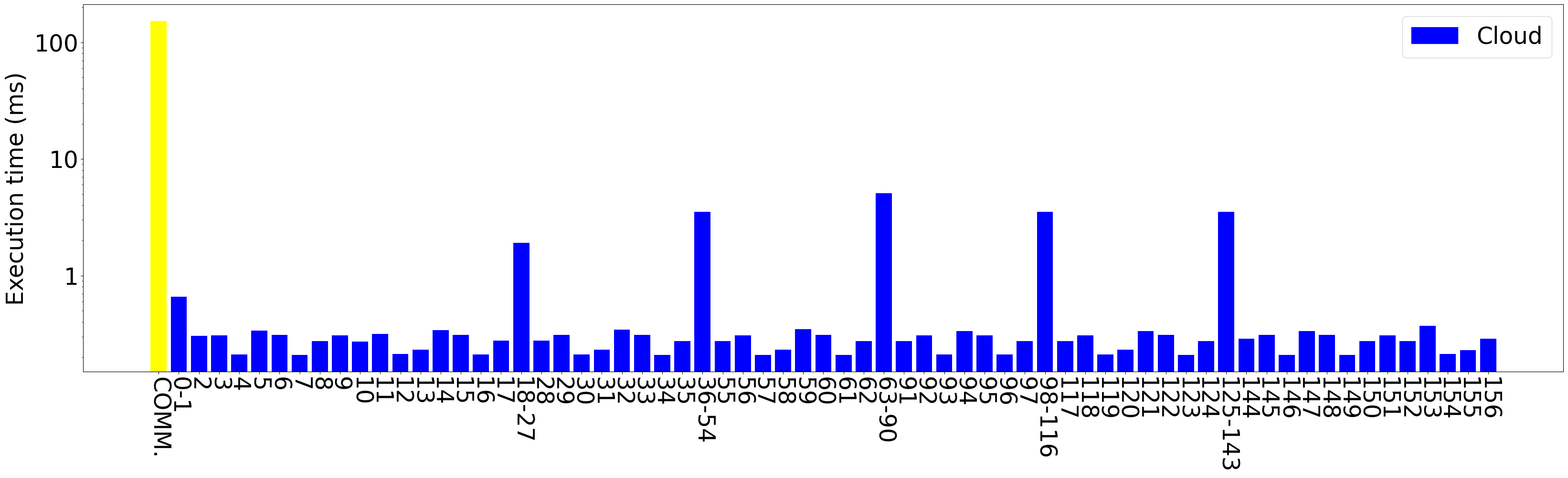}}
\end{center}
\caption{DNN partition of MobileNetV2 with lowest end-to-end latency for different network conditions}
\label{fig:network-moblilenetv2}
\end{figure}

However, Figure~\ref{fig:network-moblilenetv2} demonstrates the end-to-end latency of MobileNetV2 (that has sub-second execution performance when it is device-native) under 3G and 4G conditions. In the 3G context, the DNN has the least inference time when the DNN is device-native. However, in the 4G context, given a lower latency network, the DNN is performance efficient when it is cloud-native. 
The above highlights the capability of \texttt{Scission} to identify optimal DNN partitions under different network conditions. 

\subsubsection{DNN partitioning under different input data sizes}
\begin{figure}[ht]
	\centering
	\includegraphics[width=0.44\textwidth]{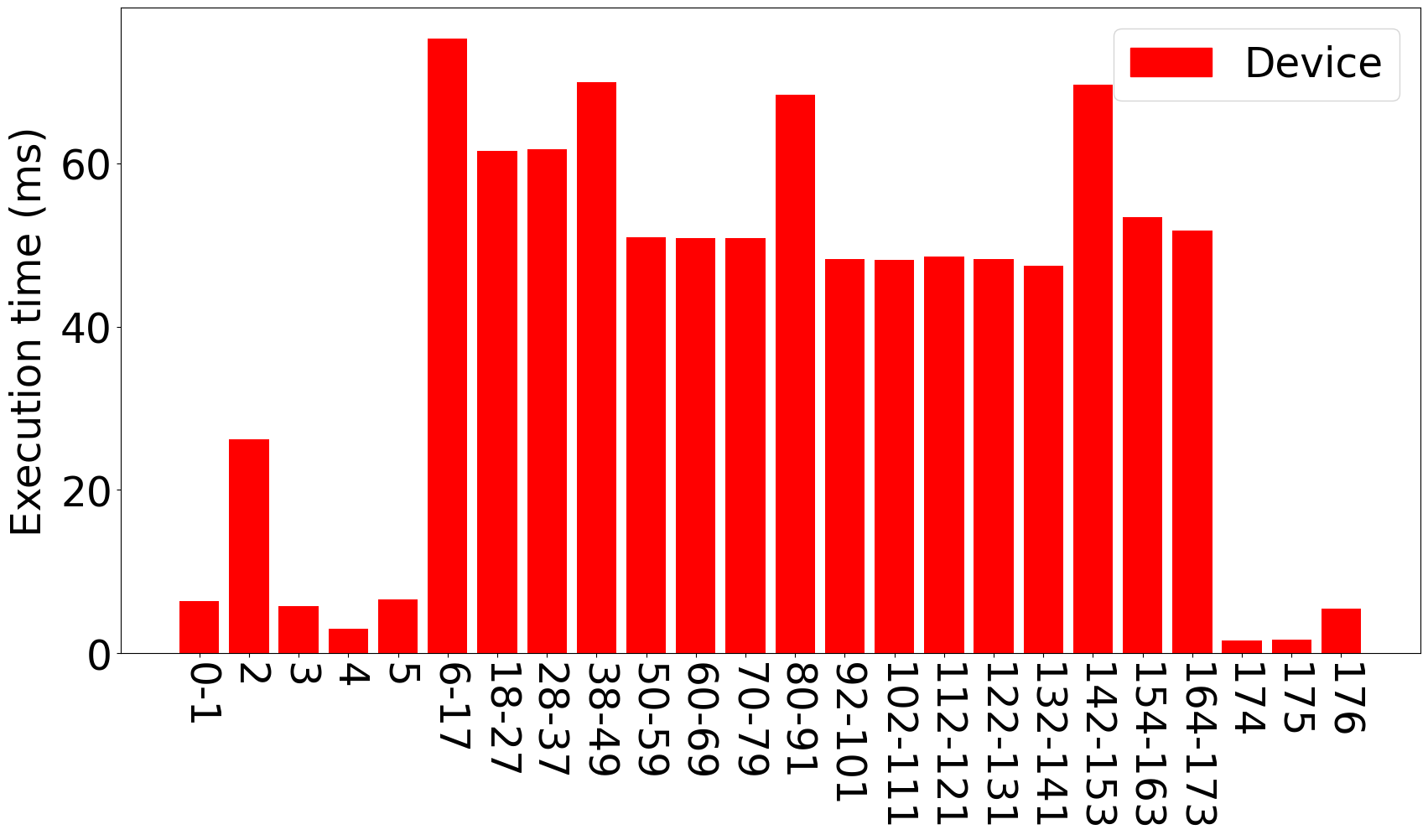}
	\caption{DNN partition of ResNet50 with lowest end-to-end latency in a 3G network when input data size is 170KB (instead of 150KB).}
\label{fig:3g-best-resnet50-170}
\end{figure}

If the input image size were increased from 150KB to 170KB, then for ResNet50 under 3G conditions, a device-native execution is determined by \texttt{Scission} to be performance efficient as shown in Figure~\ref{fig:3g-best-resnet50-170}. This is in contrast to a cloud-native execution that \texttt{Scission} identifies as performance efficient for a 150KB input image size (Figure~\ref{fig:3g-best-resnet50}).

\subsubsection{DNN partitioning under user-defined constraints}
\begin{figure}[ht]
\begin{center}
	\subfloat[3G]
	{\label{fig:3g-all-vgg19}
	\includegraphics[width=0.233\textwidth]{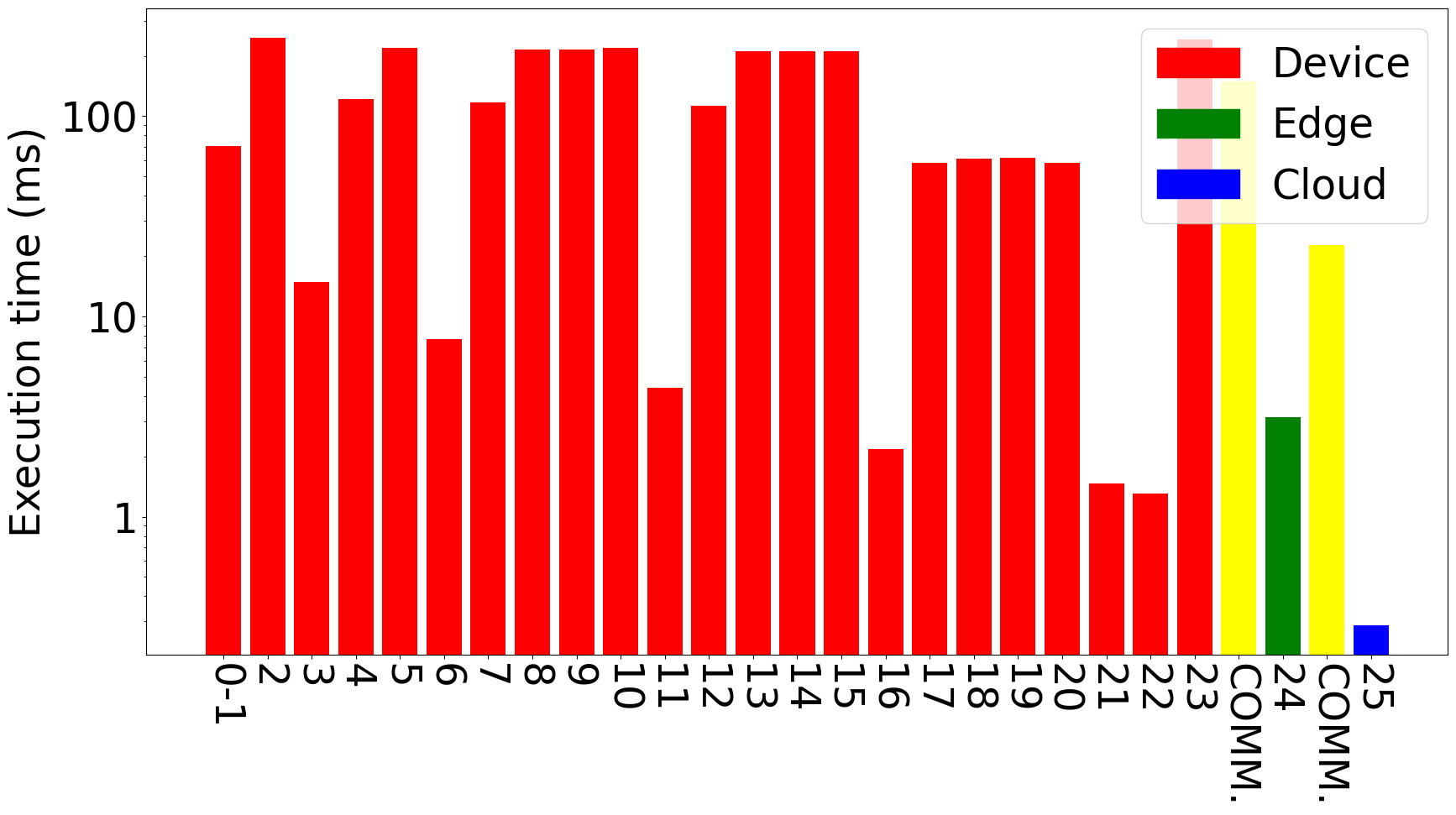}}
\hfill
	\subfloat[4G]
	{\label{fig:4g-all-vgg19}
	\includegraphics[width=0.233\textwidth]{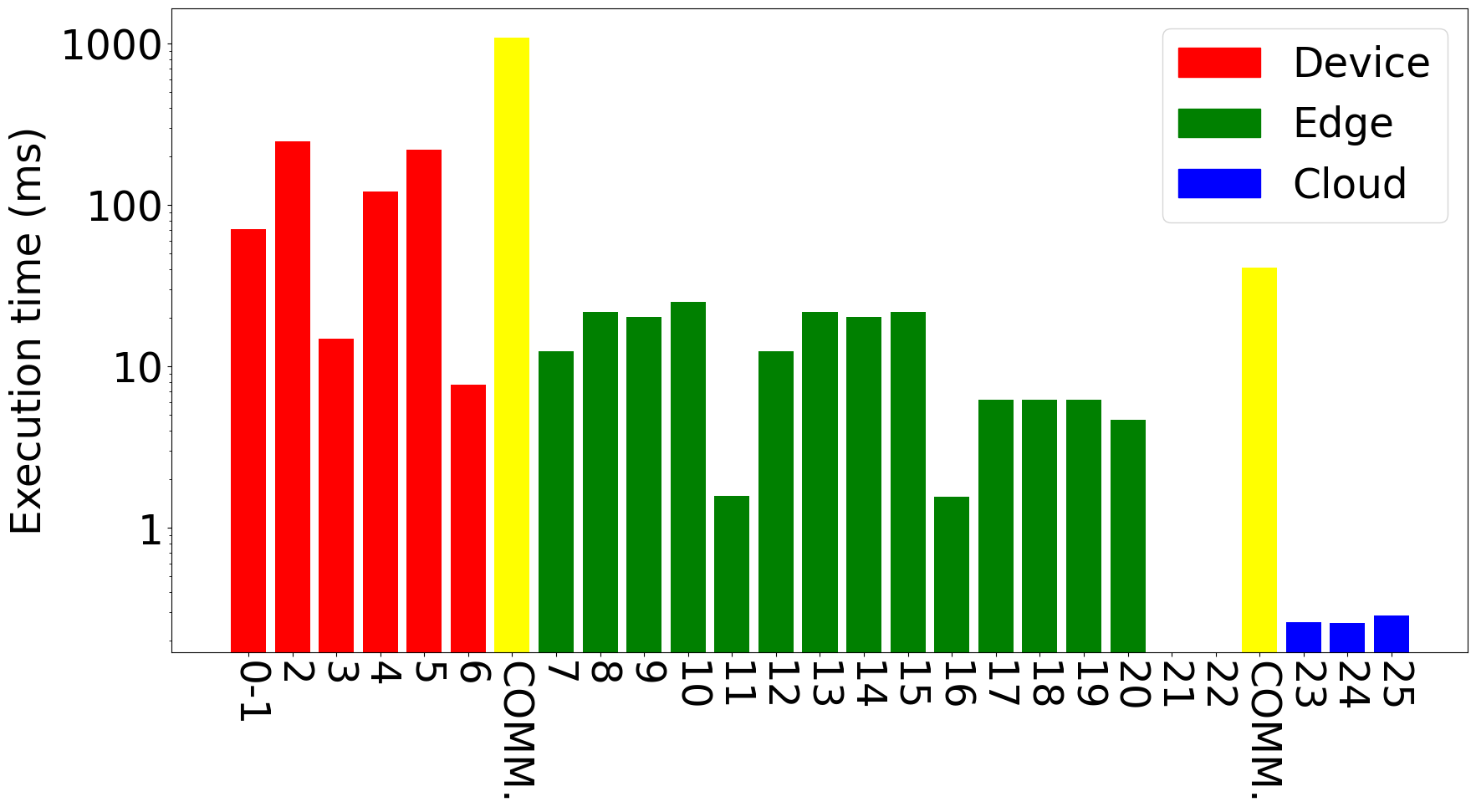}}
\end{center}
\caption{DNN partition of VGG19 with lowest end-to-end latency when the constraint imposed is that the device, edge and cloud must be used in different networks.}
\label{fig:vgg19-all}
\end{figure}

\begin{figure}[ht]
\begin{center}
	\subfloat[3G]
	{\label{fig:3g-all-resnet50}
	\includegraphics[width=0.233\textwidth]{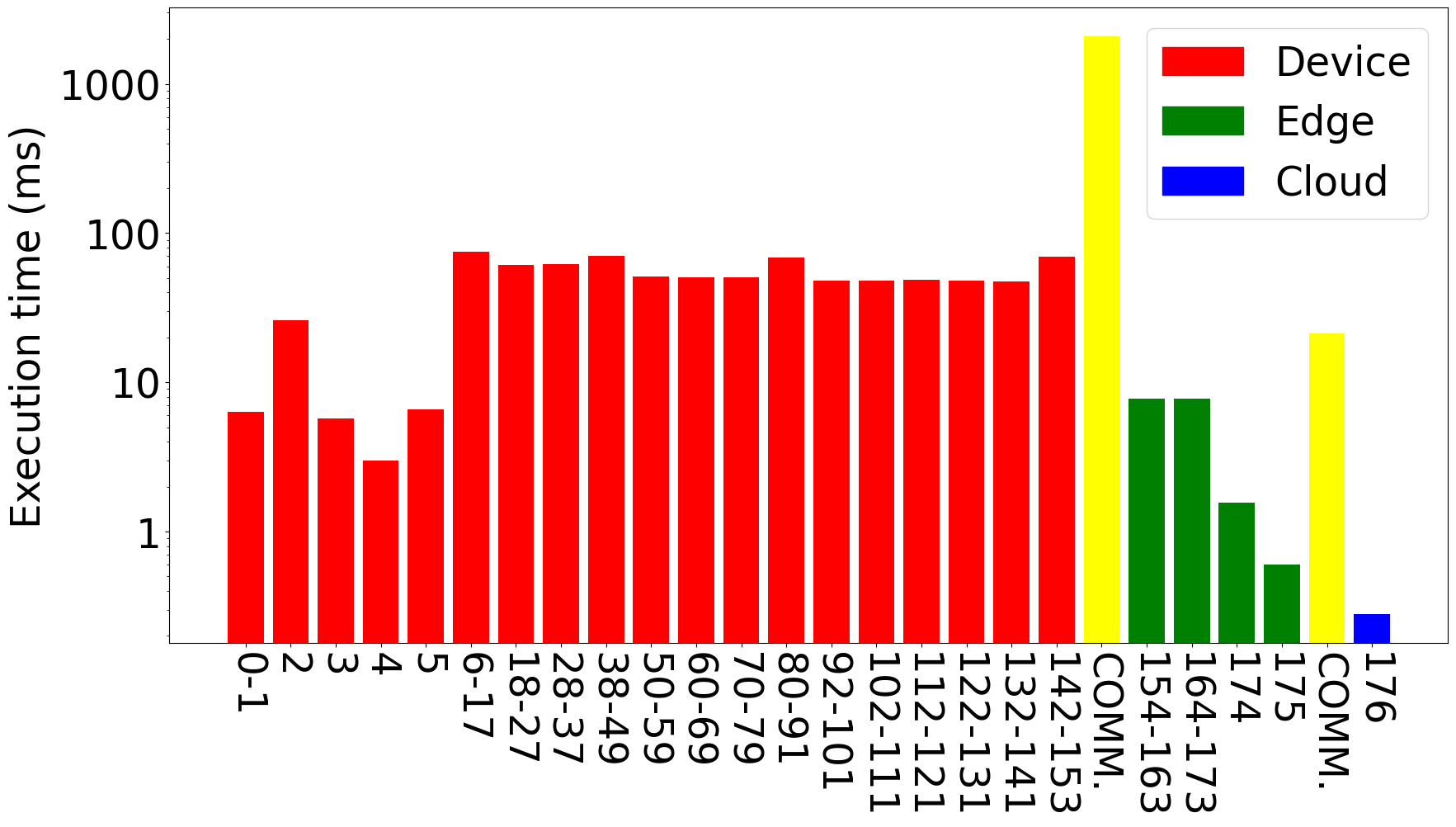}}
\hfill
	\subfloat[4G]
	{\label{fig:4g-all-resnet50}
	\includegraphics[width=0.233\textwidth]{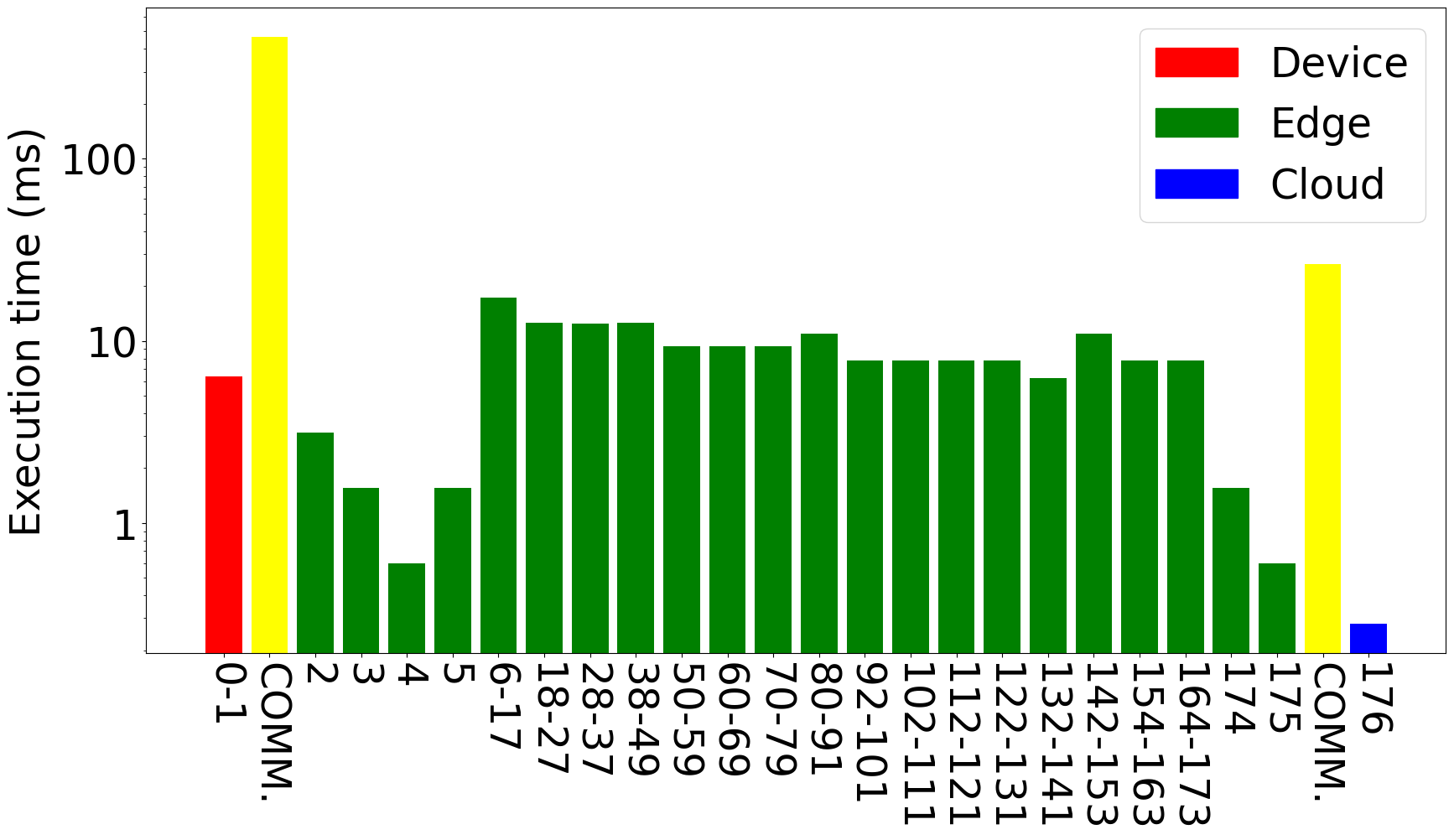}}
\end{center}
\caption{DNN partition of ResNet50 with lowest end-to-end latency when the constraint imposed is that the device, edge and cloud must be used in different networks.}
\label{fig:resnet50-all}
\end{figure}

Figure~\ref{fig:vgg19-all} and Figure~\ref{fig:resnet50-all} are exemplars of performance efficient distributed execution of the DNN when the constraint imposed is that the entire resource pipeline must be employed. The results are shown for 3G and 4G network conditions for VGG19 and ResNet50. The difference in the optimal DNN partition is immediately evident. For example, the optimal partition configuration for VGG19 in a 3G network is: device executes Layers 0-23, edge executes Layer 24 and cloud executes Layer 25 (refer Figure~\ref{fig:3g-all-vgg19}). However, in a 4G network, the optimal partition configuration is: device executes Layers 0-6, edge executes Layers 7-22, and cloud executes Layers 23-25 (refer Figure~\ref{fig:4g-all-vgg19}).

\subsubsection{DNN partitioning for comparing different target hardware resource pipelines}
Two examples from InceptionV3 and DenseNet169 highlight that \texttt{Scission} can compare target hardware resource pipelines specified by a user for identifying which resource pipeline is performance efficient. 

\begin{figure}[ht]
\begin{center}
	\subfloat[Edge (1)]
	{\label{fig:vm2Core-best-inceptionv3}
	\includegraphics[width=0.233\textwidth]{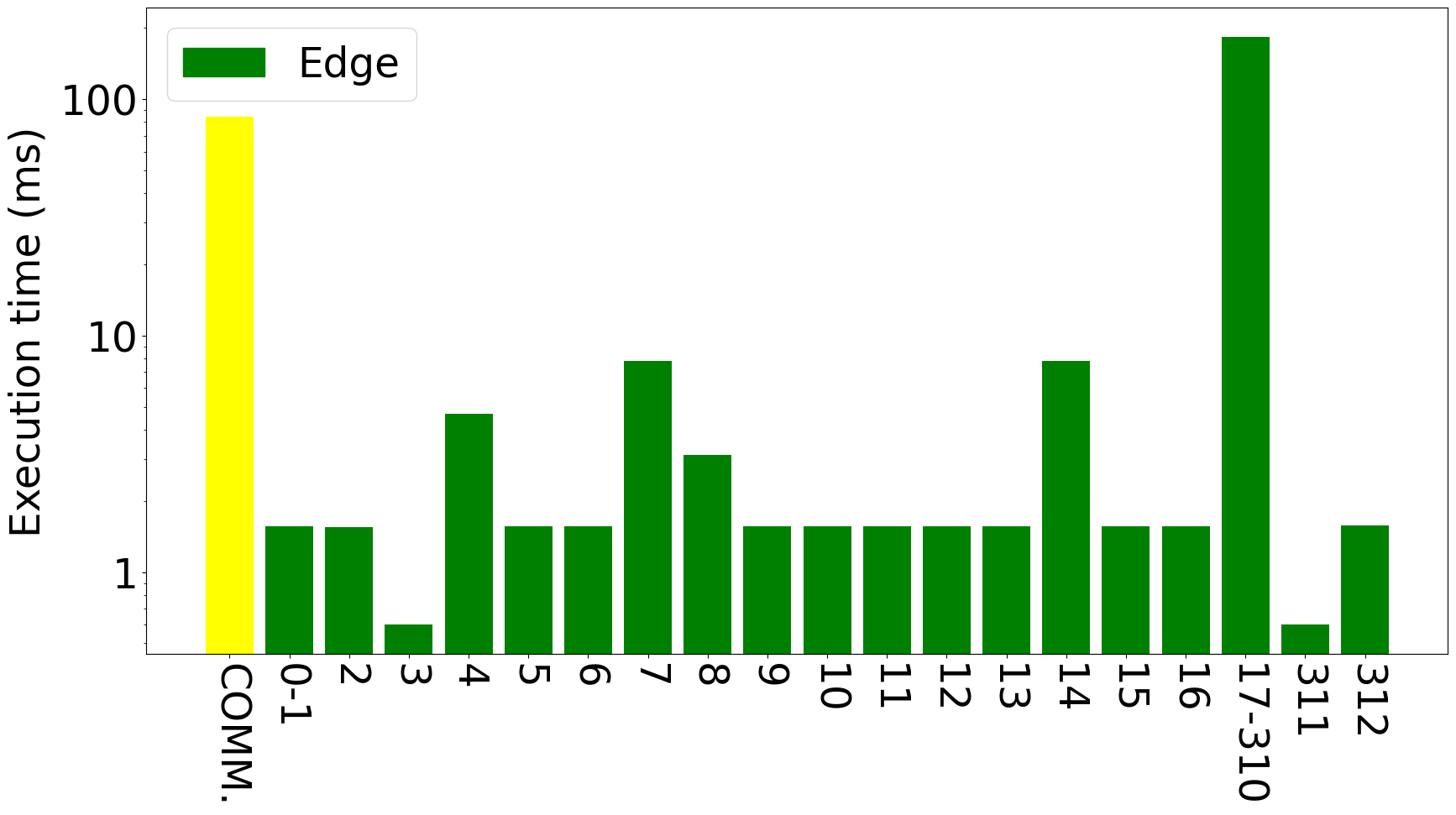}}
\hfill
	\subfloat[Edge (2)]
	{\label{fig:2500-best-incetionv3}
	\includegraphics[width=0.233\textwidth]{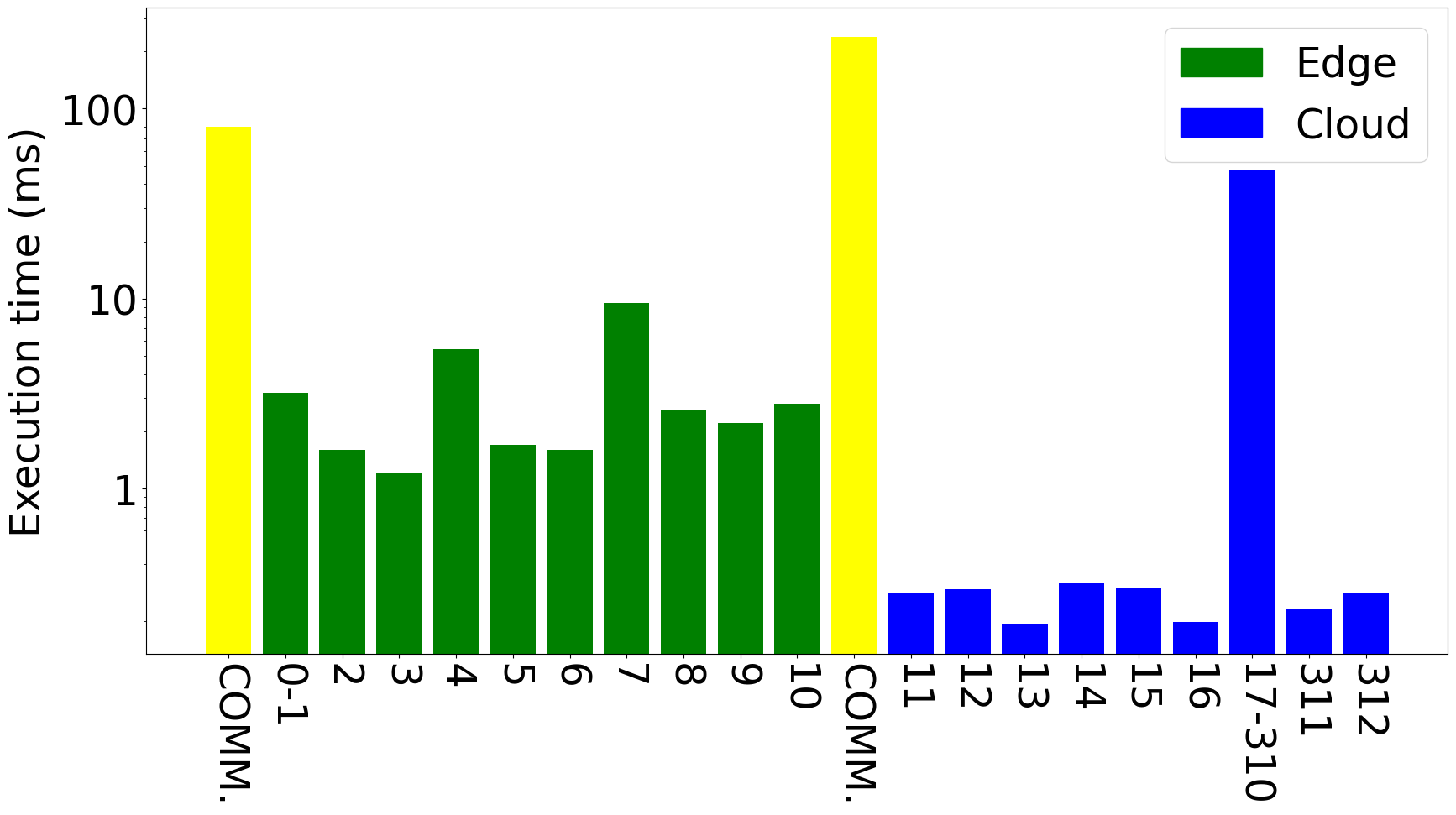}}
\end{center}
\caption{Lowest latency executions of InceptionV3 when the edge must be used in a wired network with the device.}
\label{fig:incpetionv3-edgecomparison}
\end{figure}

Figure~\ref{fig:incpetionv3-edgecomparison} considers the execution of InceptionV3 when the edge resource must be used in a pipeline and the device is connected to the edge resource via a wired connection for two different edge resources. Although the edge-native execution of InceptionV3 on Edge (1) and Edge (2) differed only by 0.07 seconds, the DNN partition configuration when the resource pipeline has Edge (1) and Edge (2) is different. The DNN partition is sensitive to the hardware capabilities of different resources in the pipeline. Since these are subtle, it would not be evident to a human, and therefore demonstrates the value of a tool, such as \texttt{Scission}.

\begin{figure}[ht]
\begin{center}
	\subfloat[Edge (1)]
	{\label{fig:vm2Core-best-inceptionv3-full}
	\includegraphics[width=0.233\textwidth]{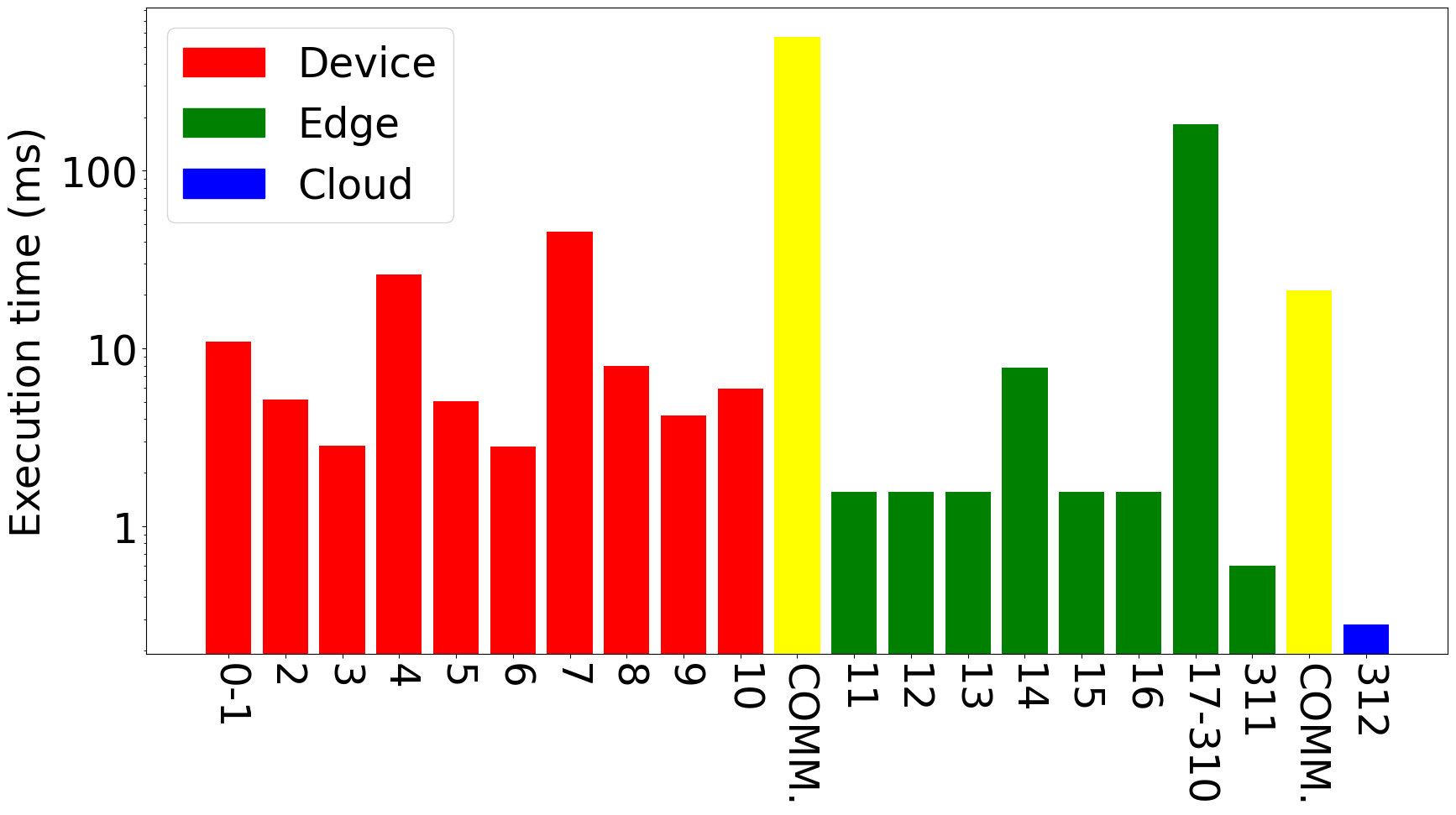}}
\hfill
	\subfloat[Edge (2)]
	{\label{fig:2500-best-incetionv3-full}
	\includegraphics[width=0.233\textwidth]{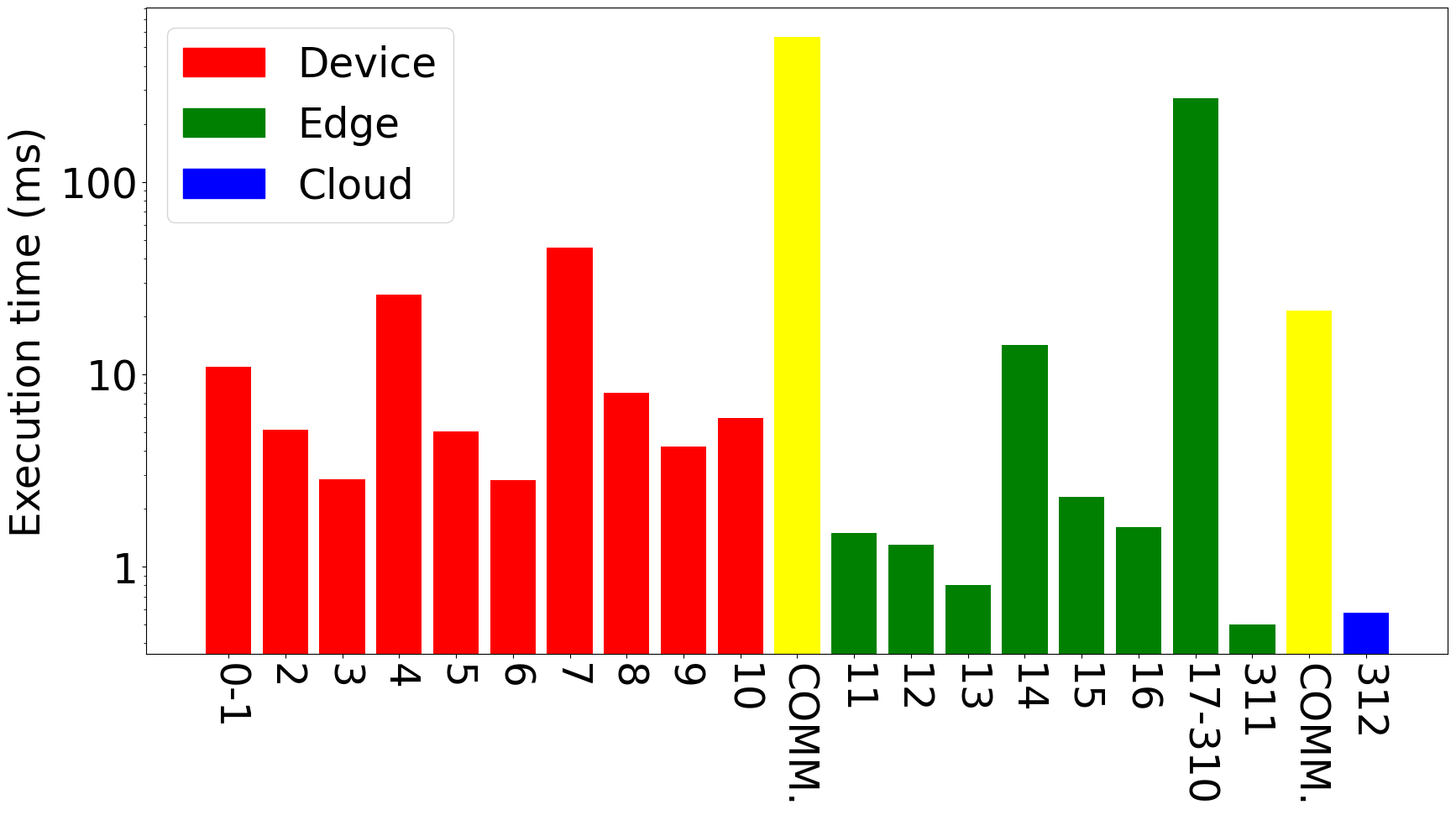}}
\end{center}
\caption{Lowest latency executions of InceptionV3 when the device, edge and cloud (entire resource pipeline) must be used in a wired network with the device.}
\label{fig:incpetionv3-edgecomparison-full}
\end{figure}

\begin{figure}[ht]
\begin{center}
	\subfloat[Edge (1)]
	{\label{fig:vm2Core-best-densenet169}
	\includegraphics[width=0.233\textwidth]{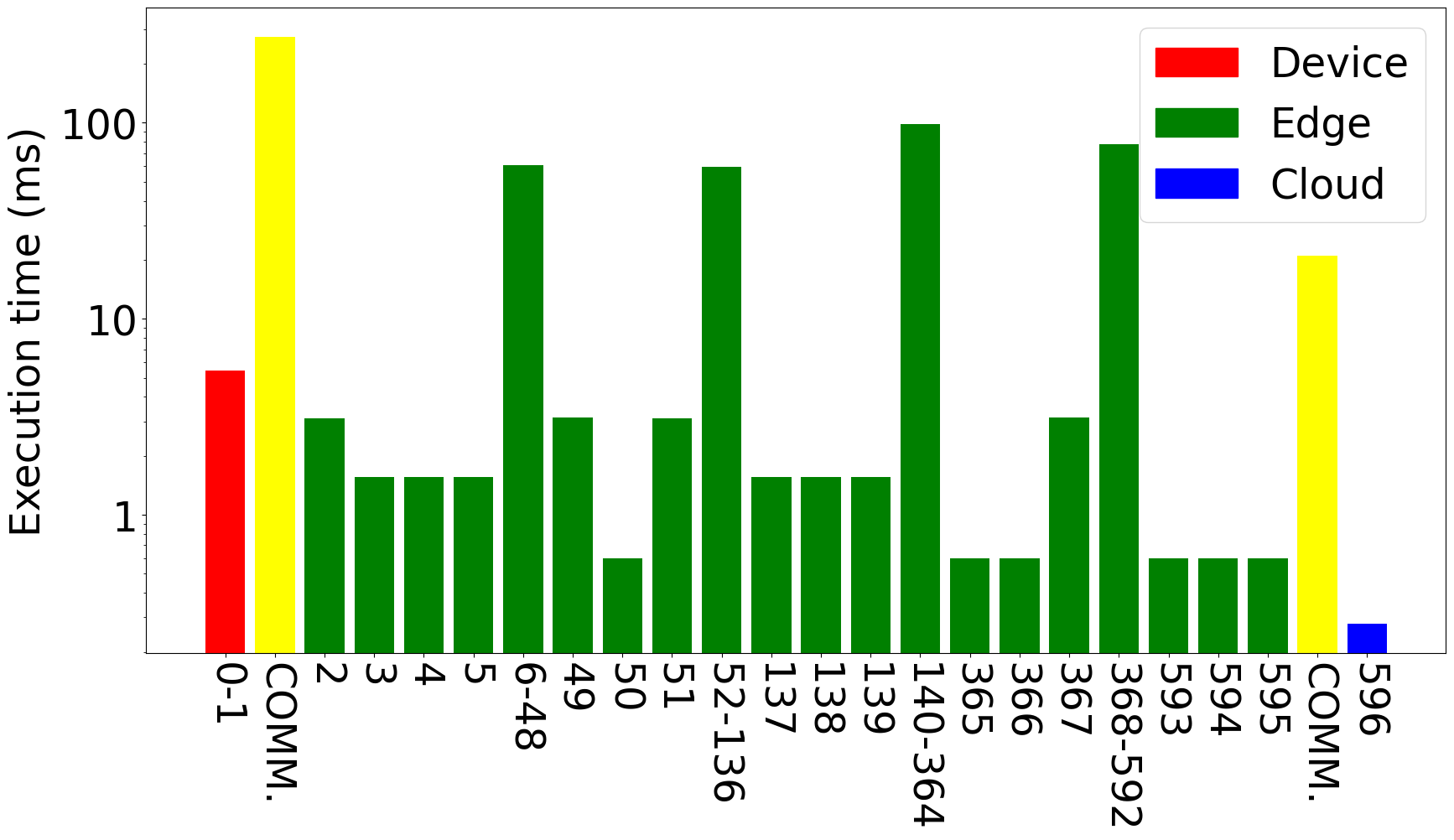}}
\hfill
	\subfloat[Edge (2)]
	{\label{fig:2500-best-densenet169}
	\includegraphics[width=0.233\textwidth]{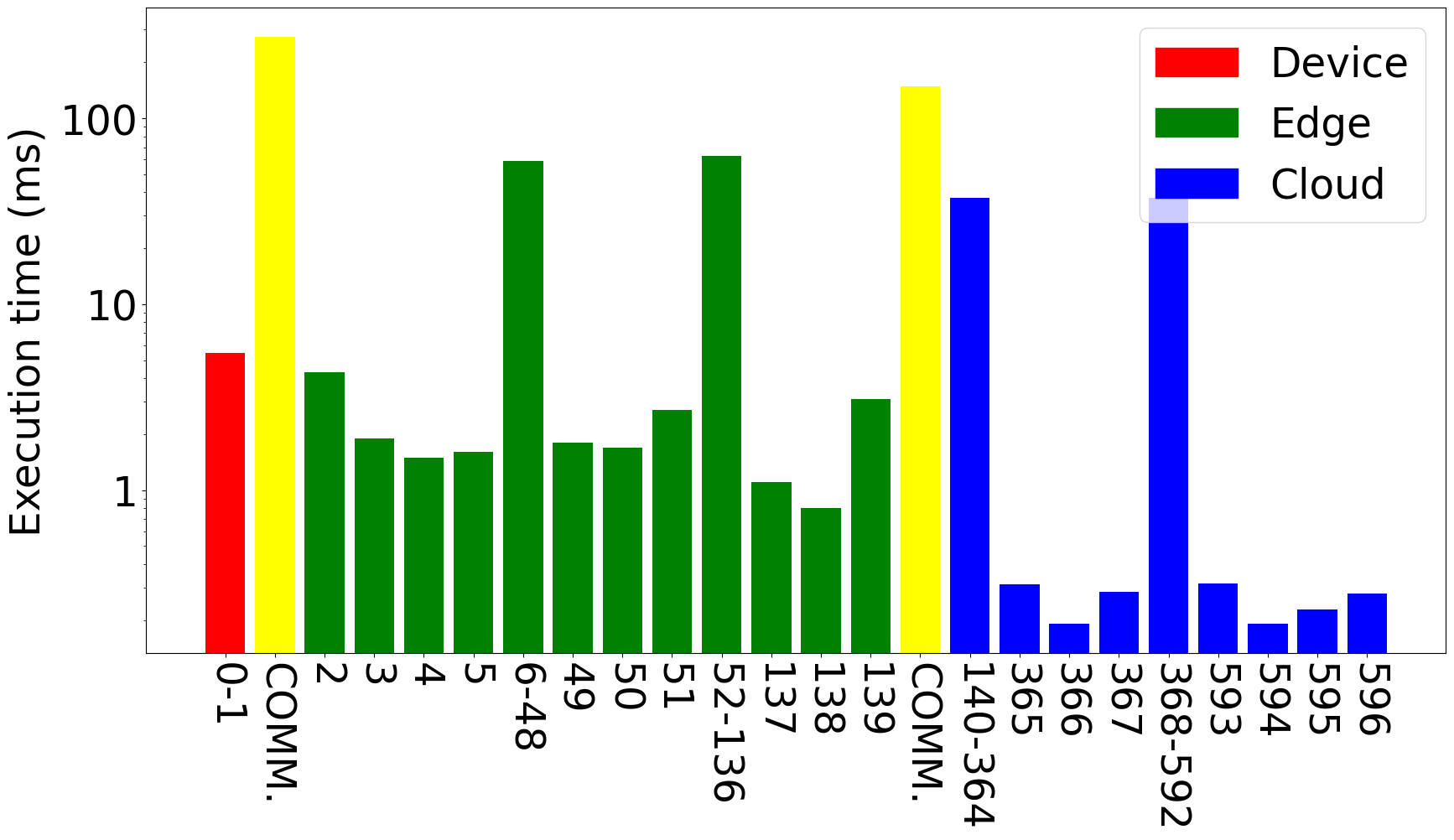}}
\end{center}
\caption{Lowest latency executions of DenseNet169 when the device, edge and cloud (entire resource pipeline) must be used in a wired network with the device.}
\label{fig:densenet169-edgecomparison}
\end{figure}

Figure~\ref{fig:incpetionv3-edgecomparison-full} and Figure~\ref{fig:densenet169-edgecomparison} considers the distributed execution of InceptionV3 and DenseNet169, respectively, for the entire resource pipeline (device, edge and cloud) when a device is connected to the edge through a wired connection for two different edge resources. 
For the same resource pipelines it is noted that for InceptionV3 there is no change to the partition configuration whereas for DenseNet169 the partition configuration changes. 

\subsubsection{Top $N$ performance-efficient DNN partitions}
\texttt{Scission} provides a list of potential candidate DNN partitions. 
Table~\ref{table:top3} shows the top three partitions with lowest end-to-end latency of ResNet50 for four different distributed pipelines that use a wired network between the device and the edge. 

\begin{figure}[!htp]
\begin{center}
	\subfloat[First rank]
	{\label{fig:vm2Core-best-resnet50}
	\includegraphics[width=0.233\textwidth]{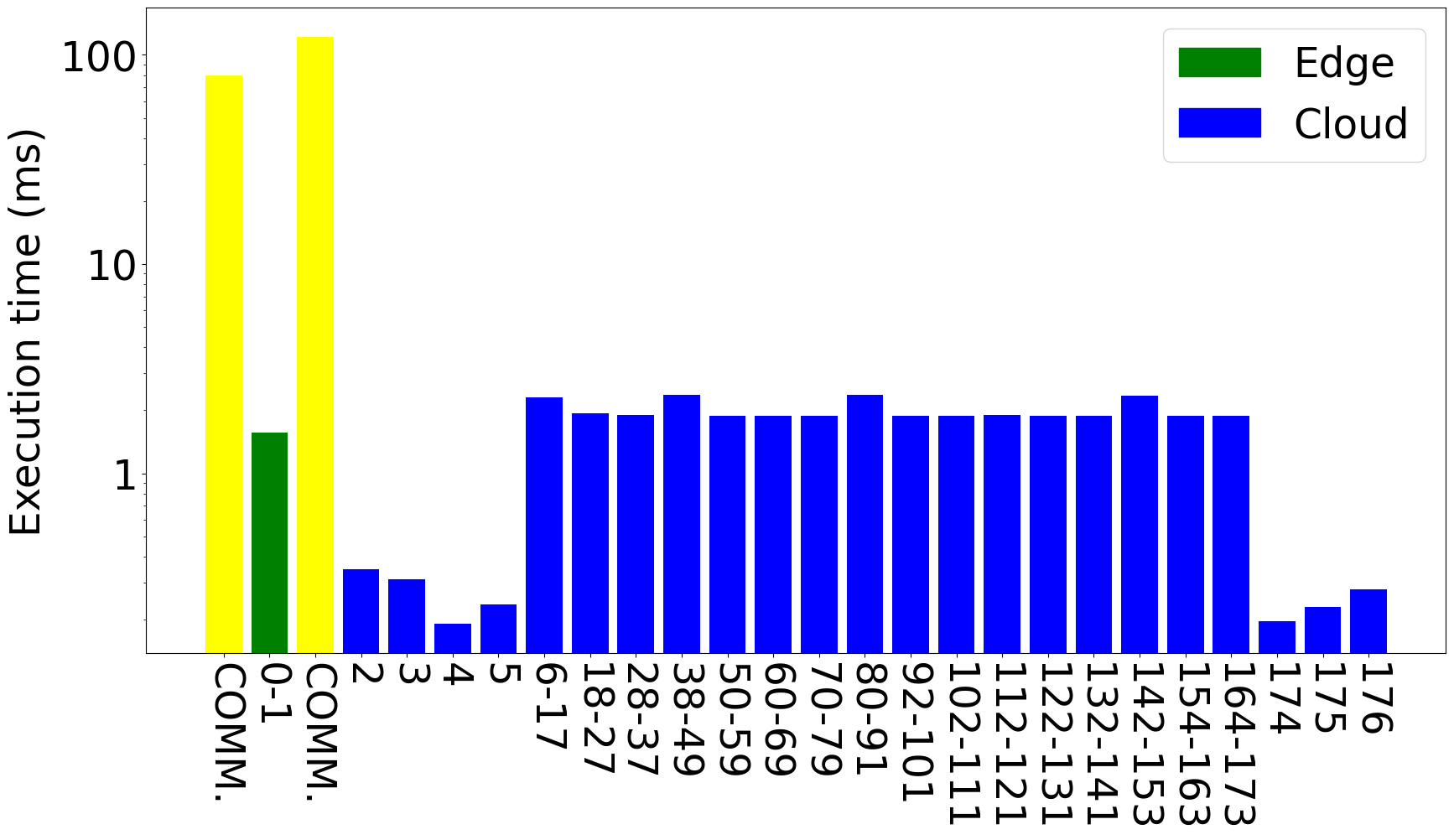}}
\hfill
	\subfloat[Second rank]
	{\label{fig:vm2Core-2nd-resnet50}
	\includegraphics[width=0.233\textwidth]{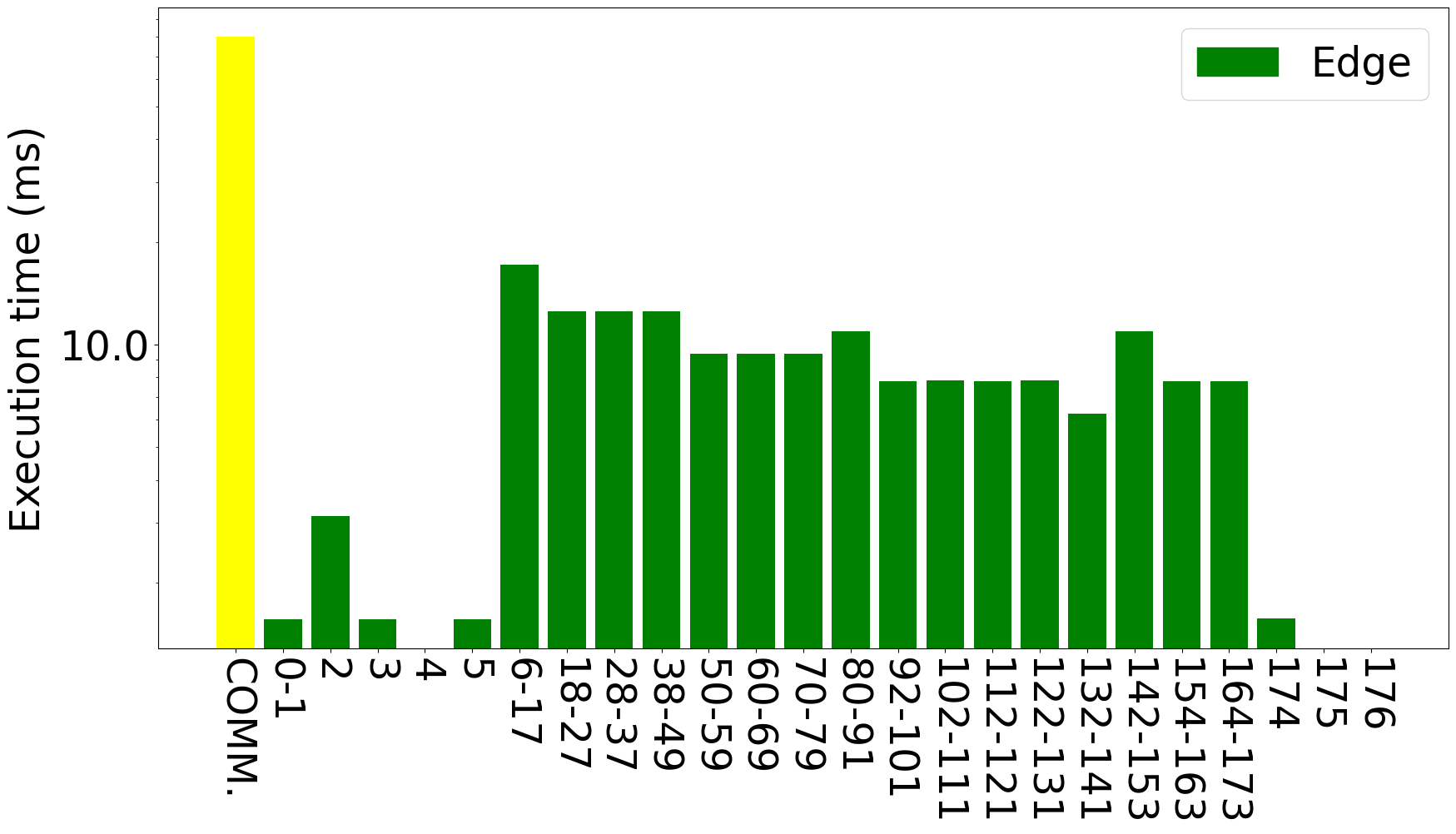}}
\end{center}
\caption{Top two DNN partitions with lowest end-to-end latency for ResNet50 when Edge (1) must be used and the device is connected to the edge via a wired network.}
\label{fig:resnet50-firstandsecond}
\end{figure}

Figure~\ref{fig:resnet50-firstandsecond} shows the DNN partitions with the first and second lowest end-to-end latencies for ResNet50 when Edge(1) must be used in the resource pipeline. The fastest partition requires offloading the most layers to the cloud, resulting in an end-to-end latency of 0.237 seconds, transferring a total of 0.785MB across the resources. On the other hand, the second DNN partition is an edge-only execution that has an end-to-end latency of 0.248 seconds, and only requires the input 150KB to be transferred to the edge. The benefit of the second partition is that it uses a much lower bandwidth than the first partition. 
\texttt{Scission} thus provides a user with a list of potential DNN configurations each of which might benefit in different scenarios.

\begin{table}
\centering
\caption{Top 3 DNN partitions with the lowest end-to-end latency for ResNet50 across different distributed resource pipelines.}
\label{table:top3}
\begin{tabular}{@{}|l|l|l|c|c|@{}}
\hline
\multicolumn{3}{|c|}{\textbf{Layers}}                                            & \multirow{2}{*}{\begin{tabular}[c]{@{}l@{}}\textbf{End-to-end}\\\textbf{latency (s)}\end{tabular}} & \multirow{2}{*}{\begin{tabular}[c]{@{}l@{}}\textbf{Total data}\\\textbf{transfer (MB)}\end{tabular}} \\ \cline{1-3}
\textbf{Device}                & \textbf{Edge (1)}                  & \textbf{Cloud (GPU)}                 &                                                                                    &                                           \\ \hline
\multicolumn{5}{|c|}{\textit{Device-Edge pipeline}}                                                                                                                                                               \\ \hline
0-1                   & 2-176                 & \multicolumn{1}{c|}{-} & 0.446                                                                              & 0.634                                     \\
0-91                  & 92-176                & \multicolumn{1}{c|}{-} & 0.944                                                                              & 0.831                                      \\
0-175                 & 176                   & \multicolumn{1}{c|}{-} & 0.979                                                                              & 0.008                                     \\ \hline
\multicolumn{5}{|c|}{\textit{Device-Cloud pipeline}}                                                                                                                                                              \\ \hline
0-1                   & \multicolumn{1}{c|}{-} & 2-176                 & 0.339                                                                              & 0.635                                     \\
0-91                  & \multicolumn{1}{c|}{-} & 92-176                & 0.920                                                                              & 0.803                                     \\
0-101                 & \multicolumn{1}{c|}{-} & 102-176               & 0.996                                                                              & 0.803                                     \\ \hline
\multicolumn{5}{|c|}{\textit{Edge-Cloud pipeline}}                                                                                                                                                                \\ \hline
\multicolumn{1}{|c|}{-} & 0-1                   & 2-176                 & 0.237                                                                              & 0.785                                     \\
\multicolumn{1}{|c|}{-} & 0-175                 & 176                   & 0.269                                                                              & 0.159                                     \\
\multicolumn{1}{|c|}{-} & 0-153                 & 154-176               & 0.319                                                                              & 0.552                                     \\ \hline
\multicolumn{5}{|c|}{\textit{Device-Edge-Cloud pipeline}}                                                                                                                                                         \\ \hline
0-1                   & 2-175                 & 176                   & 0.468                                                                              & 0.643                                     \\
0-1                   & 2-153                 & 154-176               & 0.517                                                                              & 1.036                                     \\
0-1                   & 2-163                 & 164-176               & 0.523                                                                              & 1.036                                     \\ \hline
\end{tabular}
\end{table}

\subsection{Summary}
The following are observations from the results:

1) \texttt{Scission} takes between 1.52 and 38.14 seconds for generating the benchmark data for different DNNs on the cloud and edge. Due to this overhead \texttt{Scission} would be more appropriate for responding to operational changes periodically rather than in real-time. 

2) DNN partitioning is affected by different network conditions. Although a cloud-native DNN execution was beneficial for some of the examples (VGG19 and ResNet50), it was noted that MobileNetV2 presented the possibility of both a device-native and cloud-native execution for 3G and 4G networks respectively.

3) A slightly larger input data of 170KB over 150KB changes the DNN partition of ResNet50. This highlights the potential sensitivity of DNN partitioning to data sizes. These are subtle and not quickly evident to manual inspection. 

4) User constraints, such as requiring the use of the entire resource pipeline, affects DNN partitions. The sequence of layers on the device, edge and cloud change for different networks, such as VGG19 and ResNet50. These cannot manually be identified. 

5) Variation in the edge hardware characteristics affects DNN partitioning. For InceptionV3 it was noted that using two different edge resources did not result in different partition configurations. However, the difference in performance between the two edge resources resulted in different configurations for Densenet169.

6) Obtaining a set of ranked configurations can help maximize performance in different scenarios. For example, the fastest partition with the lowest end-to-end latency for ResNet50 when a certain edge resource is to be utilized has more layers on the cloud. The second fastest partition is edge-native (suitable for enhanced privacy).

The results that can be observed on \texttt{Scission} are exhaustive. The above is only a subset of those observations arising from the experimental results. The need for such a tool is essential as more complex DNNs are appearing and is required to optimally leverage the edge and maximize the performance of distributed DNNs.   


\section{Related Work}
\label{sec:relatedwork}
Many applications have been demonstrated to benefit from using the edge either by running them natively on the edge, across the cloud and edge, or across the cloud, edge and device~\cite{intro-07, intro-08, intro-09}. 
DNNs are an example application that can be executed natively on a single resource, such as a end user device, or on the edge or cloud, or in a distributed manner across multiple resources~\cite{edge-dnn-survey-1}.    
DNN partitioning is one approach that is essential for the distributed execution of DNNs~\cite{edge-dnn-survey-1, edge-dnn-survey-2}. This has gained prominence with the upcoming paradigms in distributed systems, such as edge computing~\cite{edgecomp-01, edgecomp-02, edgecomp-03}, because by using an edge resource a series of layers of the DNN can be executed closer to the input data source, thereby reducing the ingress bandwidth demands and end-to-end latency in a resource rich environment.

There are two main methodologies that have been considered in DNN partitioning for inference (DNN partitioning for training is not considered in this paper). The first is DNN layer distribution and the second is DNN sub-model distribution. \textit{DNN layer distribution} refers to the distribution of a sequence of layers on to a resource by assuming that the resource has access to the entire pre-trained model and weights~\cite{neurosurgeon, jalad, dynamicdnnsurgery} (this methodology will be further considered). 

\textit{DNN sub-model distribution} on the other hand refers to slicing the DNN model for different resources and does not require the entire model, rather it only requires the metadata relevant to the slice of the model being executed~\cite{ionn, deepx}. IONN introduces the concept of incremental offloading in which a DNN is partitioned and incrementally uploaded on to an edge server so as to enable partial execution of the DNN even before the entire DNN is available on the edge server~\cite{ionn}. DeepX partitions the DNN model into several sub-models, which are then distributed to the edge~\cite{deepx}.

However, DNN layer distribution is a less intrusive method than DNN sub-model distribution as it does not require the DNN to be modified. Regardless, both methodologies require the identification of valid and optimal partitioning points for deploying optimal DNN partitions across resources, given the numerous combinations that may be possible. \texttt{Scission} is positioned as a tool to be used by system and network administrators for maximizing distributed DNN performance using the edge. Therefore, the design decision is one that is less intrusive and can be broadly applied. 

Approaches adopted for determining optimal DNN partitions are: (i) Profiling and estimation-based, (ii) integer linear programming-based (ILP), (iii) structural modification-based, and (iv) benchmarking-based approaches. 

\textit{Profiling and estimation-based approaches} are popular and aim to estimate the performance against metrics, such as end-to-end latency, energy or a combination, for each layer type in the DNN. Four examples of this approach are presented. 

Neurosurgeon is one example in which a regression-based method is used for estimating optimal partitions between a device and the cloud~\cite{neurosurgeon}. This is achieved by building models on the performance of individual types of layers and their configuration. 
DeepWear similarly uses a similar approach to train prediction models to estimate latency and energy consumption of four popular layer types and their parameter combinations across a wearable and its paired device~\cite{deepwear}. The models are also trained with device-specific latencies and energy prediction models. Musical Chair is another profiling and estimation-based approach that develops behavioral models that are trained to estimate the latency and memory usage of specific layer configurations~\cite{musicalchair}. 
Couper is another such approach in which the end-to-end latency of each potential partition is verified on a set of resources and then assumed as a direct correlation to hardware capability for other resource configurations~\cite{couper}. 
These approaches generally work well within the space they are trained for. However, if a new layer type/configuration or hardware emerges, then the estimation models will not be accurate. In addition, many of these approaches make assumptions regarding the execution behavior of different layers on the underlying hardware. It is not entirely possible to accurately model the execution profiles on complex hardware architectures. 

\textit{ILP-based approaches} have also been considered for DNN partitioning. Within the context, the partitioning problem is formulated as an ILP problem with the aim to find an optimal partition that minimizes the inference latency and maximizes accuracy~\cite{jalad, jointdnn}. ILP techniques can be time consuming. 

\textit{Structural modification-based approaches} can efficiently partition DNNs, but in an intrusive manner. It can be achieved realistically only by modifying underlying libraries of existing frameworks or by writing bespoke code for DNNs. However, these approaches provide a fine-grained control over DNN partitioning.  
Examples include DeepThings~\cite{deepthings} and MoDNN~\cite{modnn}. DeepThings utilizes fuse tile partitioning, in which a DNN is not partitioned horizontally (based on layers), rather they are partitioned vertically to reduce resource footprint~\cite{deepthings}. 
MoDNN is developed for distributing DNNs across different nodes of the same cluster~\cite{modnn}. Three approaches are presented: (i) for partitioning the convolutional layers, biased one-dimensional partitioning, (ii) for partitioning the weights, modified spectral co-clustering (the fully connected layers are dependent on weights), and (iii) for partitioning sparse fully connected layers, fine-grain cross partition are proposed. 

\textit{Benchmarking-based approaches} are proposed so that actual measurements or observations are made on the target hardware resource. No assumptions are made of the underlying hardware or performance of the layers on the hardware and therefore are more reliable. In these approaches, benchmarking data of the DNN on the hardware is first obtained. Then during deployment, a snapshot of the operational environment (for example, load on the network and compute resource) is taken and the optimal partition is calculated. This approach is minimally intrusive, requires no modification to the code, and is a pragmatic solution in the complex space of DNNs with many layers (and layer types and configurations) and the availability of diverse hardware resources, although they cannot operate in real-time and can only be used periodically. \texttt{Scission} proposed in this paper is therefore positioned as a benchmarking-based approach. This approach is used by LAVEA for distributed video analytics~\cite{lavea}. 

Other approaches, such as approximation-based are also considered in the literature~\cite{dynamicdnnsurgery}, but are not considered here.

\section{Conclusions}
\label{sec:conclusions}
This paper presented \texttt{Scission}, a tool for automated benchmarking of DNNs on a given set of target device, edge and cloud resources for determining the optimal partition for maximizing DNN performance. \texttt{Scission} is underpinned by a benchmarking approach that determines the combination of potential target hardware resources and the sequence of layers that should be distributed for maximizing distributed DNN performance while accounting for user-defined objectives. 
\texttt{Scission} relies on empirical data and does not estimate performance by making assumptions of the target hardware or the DNN layers. 
Experimental studies were carried out on 18 different DNNs to demonstrate that \texttt{Scission} is a valuable tool for obtaining context-aware and performance efficient distributed DNNs. \texttt{Scission} can also make decisions that cannot be manually made by a human due to the complexity and number of dimensions affecting the search space. 

\textit{Limitations and Future Work}: 
Since \texttt{Scission} relies on exhaustive benchmarking and search it cannot be used in scenarios that need to account for rapid changes (failures or variance) given the time overhead. Nonetheless, it would prove useful in scenarios where accuracy of the partition configuration is important. Meta-heuristic optimization will be considered to rapidly respond to network congestion or resource failure. Other metrics, such as monetary costs and performance improvements, as well as trade-offs that exist among performance gain and costs, and optimal partitioning and responsiveness of the approach will be considered. The offering of \texttt{Scission} as a service will be integrated within a standard orchestration framework to monitor and partition DNNs. The current work assumes that partitioning a given DNN is beneficial and does not account for whether the performance gain may be relatively low. \texttt{Scission} can be further extended to determine whether an alternate DNN can be selected for performance gain instead of partitioning a given DNN. 

\section*{Acknowledgment}
Dr Blesson Varghese is supported by a Royal Society Short Industry Fellowship to British Telecommunications plc, UK and by funds from Rakuten Mobile, Japan. 

\balance
\bibliographystyle{IEEEtran} 
\bibliography{reference}

\end{document}